\documentclass[fleqn,twoside]{article}
\usepackage{epsfig}
\usepackage{amsfonts,amsmath}
\usepackage[headings]{espcrc2}
\readRCS
$Id: espcrc2.tex,v 1.2 2004/02/24 11:22:11 spepping Exp $
\usepackage{graphicx}
\def\half{\frac{1}{2}}

\def\e{{\epsilon}}
\def\p{\partial}

\def\r{\rightarrow}

\def\N{\mathcal{N}}
\def\M{\mathcal{M}}
\def\F{\mathcal{F}}

\def\S{\Sigma}
\def\a{\alpha}
\def\b{\beta}
\def\d{\delta}
\def\L{\Lambda}
\def\Q{\mathcal{O}}
\def\th{\theta}
\def\tp{\tilde{\psi}}
\newcommand{\bi}{\begin{itemize}}
\newcommand{\ei}{\end{itemize}}
\newcommand{\bea}{\begin{eqnarray}}
\newcommand{\eea}{\end{eqnarray}}
\newcommand{\be}{\begin{equation}}
\newcommand{\ee}{\end{equation}}
\newcommand{\non}{\nonumber}

\title{Carg\`{e}se Lectures on String Theory with Eight Supercharges}

\author{Monica Guica and Andrew Strominger \\ Jefferson Physical
Laboratory\\Harvard University \\ Cambridge, MA 02138}

\runtitle{String Theory with Eight Supercharges} \runauthor{M. Guica
and A. Strominger}

\begin{document}

\begin{abstract}
 These lectures give  an introduction
to the interrelated topics of Calabi-Yau compactification of the
type II string, black hole attractors, the all-orders entropy
formula, the dual $(0,4)$ CFT,  topological strings and the OSV
conjecture. Based on notes by MG of lectures by AS at the 2006
Carg\`{e}se summer school. \vspace{1pc}
\end{abstract}

\maketitle

\tableofcontents

\section{Introduction}
One of the best understood quantum systems is IIB string theory on
$AdS_5\times S^5$ with its dual presentation as $\N=4$ Yang Mills
gauge theory. The tractability of this system is in large part due
to the large supersymmetry group which has the maximal 32
supercharges. But a large symmetry group is both a blessing and a
curse. It is a blessing because many features of the theory can be
deduced using symmetry arguments alone. It is a curse because these
same symmetries limit the possible dynamics and questions we can
ask. For example, true tests of the various dualities which are not
already implied by the symmetries of the 32-supercharge theory are
relatively hard to find.

 As the number of supersymmetries is decreased, more dynamical
possibilities are unleashed, often with no counterparts in their
more symmetric cousins. Ultimately the most interesting and most
physically relevant case is no supersymmetry at all. But at the
moment analytic control in this case is quite limited.
\begin{figure*}[htp]
\centering
\includegraphics[height=20 cm]{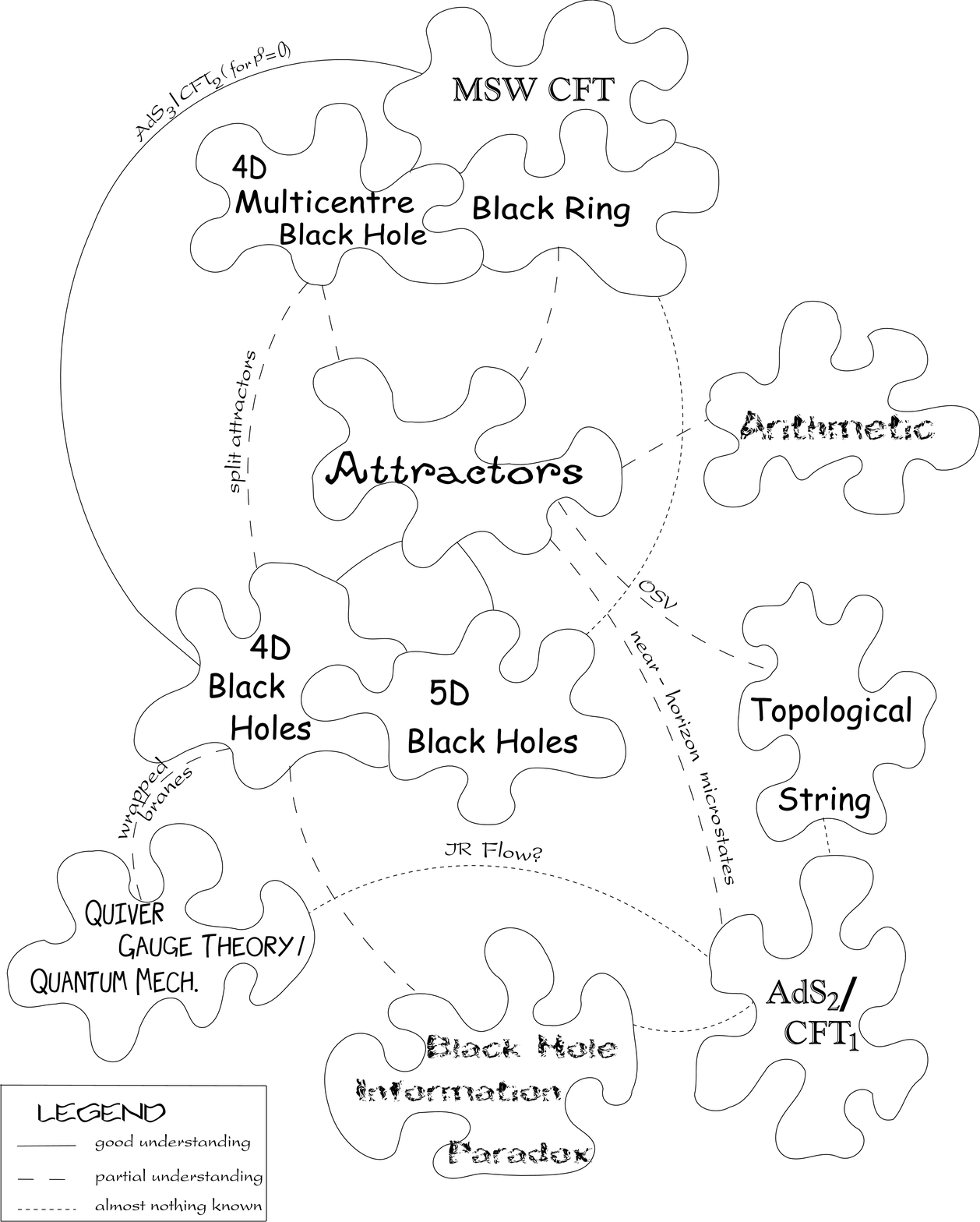}
\caption{The eight-supercharge jigsaw puzzle. It probably cannot be
assembled on the plane. }\label{puzzle}
\end{figure*}
   At the current juncture, the intermediate case of eight
   supercharges has proven to be especially fertile ground for progress.
   Eight supercharges is where black holes are first encountered\footnote{Of course, black holes
   exist in compactifications with more supersymmetry, but the near-horizon geometry
    has at most eight supersymmetries in all known examples.}, as well as Calabi-Yau spaces
and
   topological strings. Despite several decades of work in this area,
   qualitative new surprises are still appearing
   (e.g.\cite{Bates:2003vx,Elvang:2004rt,Ooguri:2004zv,Gaiotto:2005gf}). It
   is likely that a few more will still appear before the eight-supercharge
   jigsaw puzzle, illustrated in figure
   1, is pieced together. For example it has been suggested \cite{Moore:1998pn,Moore:1998zu} that
   number theory, which has yet to be fully wed to string theory,
   will play a prominent role. Another likely suspect, as yet to fully appear in
   the story,
   is the enigmatic $AdS_2/CFT_1$ duality.

   Among the fascinating connections illustrated in figure \ref{puzzle},
   these lectures will focus on those directly relevant to
   understanding the so-called OSV conjecture \cite{Ooguri:2004zv}. The agreement
   between the Bekenstein-Hawking area law and the microscopic
   state counting for certain stringy black holes with eight near-horizon
   supercharges has been the source of many deep insights into string theory and quantum gravity.
   But this area law is just the first term in an infinite series approximating
   the exact expression for black hole entropy. OSV
   conjecturally relates every term in this series to expansion
   coefficients of the topological string, thereby potentially allowing for
   ``precision tests'' of holographic duality. We are likely to have
   much to learn by precisely formulating, proving or disproving OSV.

   In these lectures we will present some introductory material on
   string theory with eight supercharges, aimed at the student
   preparing for research in this area, and building up to a discussion of the
   OSV conjecture
   in the last section.  For excellent and recent reviews at a
   more advanced level see \cite{Pioline:2006ni,Kraus:2006wn}. We begin
   in lecture 2 with the basics of Type  II Calabi-Yau
   compactification, BPS black holes, attractors, and their CFT
   duals. In lecture 3 we give a brief introduction to the
   topological string. In lecture 4 we describe the OSV conjecture
   which relates lectures 2 and 3. In lecture 5 we conclude by summarizing the
   various partition functions we have encountered along the way,
   the extent to which they are well defined, and the  relations
   between them.

\section{BPS  Calabi-Yau black holes  }

In this lecture we will briefly review the field content of type
IIA/IIB string theory compactified on a Calabi-Yau manifold, and the
resulting $\N=2$ $4d$ supergravity action. We will show that the
$4d$ theory admits supersymmetric black hole solutions of the
extremal Reissner-Nordstr\o m type, which exhibit an interesting
attractor mechanism for the vector moduli as one approaches the
horizon of the black hole. The macroscopic entropy of these black
holes is computed and it is shown how to include higher order
corrections. We briefly discuss the proposed microscopic dual, known
as the MSW CFT, for the special case of no $D6$ charge. (Finding the
dual for the general case is an important unsolved problem). We also
describe the index - the modified elliptic genus - which counts the
weighted number of BPS states in this CFT.

Throughout these lectures, we will use whichever of the type IIA/IIB
languages makes the exposition simplest. Thus the $4d$ $\N=2$ action
will be discussed mostly in the context of IIB compactifications,
while in order to talk about black holes we will switch to the IIA
description, which is also the one appropriate for understanding the
OSV conjecture.

\subsection{IIB Calabi-Yau compactification}
\medskip
The bosonic $10d$ field content of type IIB string theory is as
follows: from the NS-NS sector we get the graviton $G_{MN}$, the
antisymmetric tensor field $B_{MN}$ with field strength $H^{(3)} = d
B^{(2)}$, and the dilaton $\Phi$. From the R-R sector we have a
scalar $a$ (the axion), a two-form with $F^{(3)} = dC^{(2)}$ and a
self-dual four-form obeying (at the linearized level) $F^{(5)} =
dC^{(4)} = \star F^{(5)}$.

Compactifying six of the ten dimensions on a Calabi-Yau manifold $M$
breaks three quarters of the original 32 component supersymmetries
of type IIB supergravity. Since in four dimensions the minimal
spinors have four components, the eight supersymmetries we are left
with give $\mathcal{N}=2$ supergravity in four dimensions. This
theory - described in great detail in the fundamental paper
\cite{deWit:1984px} - has an $SU(2)_R$ $R$-symmetry under which the
two supercharges transform as doublets. The massless fields fall
into $\mathcal{N}=2$ representations labeled by their highest spins.
Three kinds of multiplets appear: \bi
\item {\bf gravity multiplet}: contains the graviton
(2), two gravitini (${3\over 2}$) in an $SU(2)_R$ doublet and the
graviphoton (1),
\item {\bf vector multiplets}: contain one photon (1), two fermions ($\half$) in an $SU(2)_R$ doublet
 and two real scalars (0),
\item {\bf hypermultiplets}: contain two hyperfermions ($\half$)
and four hyperscalars (0) in two $SU(2)_R$ doublets, \ei where in
parentheses we have written the spins of the corresponding
particles.

Now let us figure out how many multiplets of each kind we get from
the compactification\footnote{Here we will only follow what happens
to the massless bosons that arise from the compactification; the
fermions complete the supermultiplets.}. Upon dimensional reduction
from $10d$ to $4d$, the massless wave equation splits into two
pieces, schematically $ \Box_{10} = \Box_{4} + \Box_{CY} $. Harmonic
forms $\omega$ satisfy $\Box_{CY} \omega =0 $ and they are in
one-to-one correspondence with cohomology classes on $M$, whose
numbers are counted by the Hodge numbers of $M$,
$h_{p,q}$\footnote{$h_{p,q}$ is the number of harmonic forms of
antiholomorphic rank $p$ and holomorphic rank $q$. A Calabi-Yau
manifold has $h_{0,0}=1$, $h_{0,i} =0$ for $i \neq 3$, one harmonic
(0,3) form $\Omega$ and its (3,0) conjugate $\bar{\Omega}$, and
$h_{1,1} \geq 1$, $h_{1,2} \geq 0$. All other Hodge numbers are
given by Poincar\'e duality $h_{p,q} = h_{3-p,3-q}$. There is a
total of $2 h_{2,1} +2$ harmonic three-forms on $M$. }. Thus
harmonic forms on $M$ lead to massless fields in $4d$.

There are two multiplets that we obtain in $any$ compactification of
type IIB on a Calabi-Yau: the gravity multiplet and the so-called
{\it universal hypermultiplet}. The universal hypermultiplet
consists of the $10 d$ dilaton $\Phi$, the $10d$ axion $a$, plus two
more massless scalars $\chi$ and $\psi$ defined by writing $H^{(3)}
= \star \,{}_{4} d \psi$\footnote{ The Bianchi identity $d H^{(3)} =0 $
becomes $ d \star{}_4 d \psi = \Box_4 \psi =0$, which is the equation of
motion for a massless $4d$ scalar.} and $F^{(3)} = \star\,{}_4 d \chi $.
The bosonic content of the universal hypermultiplet is then $(\Phi,
a, \psi, \chi)$ and it will obviously be the same no matter which
Calabi-Yau we choose to compactify on. Importantly, the string
coupling constant $g_s = e^{\Phi}$ always belongs to this
hypermultiplet.

To figure out the remaining multiplets, choose an integral basis  of
harmonic three-forms on the Calabi-Yau,
 ($\a_{\L}, \b^{\L}$), with   $\L \in \{0, \ldots, h_{2,1} \}$, which satisfy
$$\int_M \a_{\L} \wedge \b^{\Sigma} = \d_{\L}{}^{\Sigma} \;, \;\;\;\; \int_M \a \wedge \a = \int_M \b \wedge \b =0 ,$$
define the periods as integrals of the holomorphic three-form over
the 3-cycles dual to $\a_{\L}, \b^{\L}$ as \be X^{\L} =
\int_{A_{\L}} \Omega \;, \;\;\;\;\; F_{\L} = \int_{B^{\L}} \Omega ,
\ee and let $\omega^A$ denote an integral basis of harmonic
two-forms, $A \in {1, \ldots , h_{1,1}}$. Using Greek letters for
$4d$ space-time indices and Latin letters for Calabi-Yau indices,
and $x$ for a $4d$ spacetime coordinate, the decomposition of the
various fields is then \bi
\item $G_{MN} \r$ $g_{\mu\nu}(x)$ - the $4d$ graviton; $g_{m\bar n}= i\phi_A(x) \,\omega^A_{m\bar n}$ - $h_{1,1}$ real $4d$ scalars
corresponding to K\"{a}hler deformations of the metric on $M$;
$h_{2,1}$ complex scalars $G^I$, $I=1,...h_{2,1}$ corresponding to
deformations of the complex structure of M\footnote{A complex
structure deformation mixes the holomorphic and antiholomorphic
coordinates as $z^i \r z^i + \mu^{I i}{}_{\bar j} z^{\bar j} $. If
we lower the index on $\mu^{Ii}{}_{\bar j}$ by contracting with the
holomorphic three-form $\Omega$ the resulting $G^{I}_{kl\bar
j}=\Omega_{kli} \mu^{Ii}{}_{\bar j}$ are in one-to-one
correspondence with the harmonic (2,1) forms on the Calabi-Yau.}.
The periods $X^{\L}$ provide projective coordinates on the moduli
space of complex structure deformations.
\item $B_{MN} \r \psi(x)$ -scalar in the universal hypermultiplet, as discussed;
 $B_{m\bar n} = b_A(x) \,\omega^A_{m\bar n}$ - $h_{1,1}$ real scalars.
\item $C^{(2)}_{MN} \r$ Similar to the $B_{MN}$
 case, we get one scalar field  $\chi(x)$ from the space-time part of $C^{(2)}$ and another
 $h_{1,1}$ real scalars
 $c_A(x)$ from the two forms on $M$.
\item $C^{(4)}_{MNPQ}\r$Decomposing $C^{(4)}$ as $A^{\L} (x) \a_{\L} + \tilde{A}_{\L} (x) \b^{\L}$
and imposing the self-duality condition $F^{(5)} = \star F^{(5)}$ we
get $ h_{2,1}+1 $ $4d$ massless $U(1)$ gauge bosons. We also get
$h_{2,2}=h_{1,1}$ scalars from decomposing $C^{(4)} = \gamma^A(x)
\omega_A$, where $\omega_A$ are harmonic (2,2) forms on $M$. \ei The
supersymmetry transformations (which we don't reproduce here) tell
us how these bosonic fields group into multiplets. Out of the total
of $h_{2,1}+1$ gauge fields, one linear combination -determined from
the supersymmetry transformations - has to end up in the gravity
multiplet and hence is called the graviphoton. We denote the
graviphoton by $A_{\mu}$ and the remaining gauge bosons by
$A^{I}_{\mu}$, where $I$ takes only $h_{2,1}$ values. The groupings
of the bosonic fields are then \bi
\item the gravity multiplet ($g_{\mu\nu}$, $A_{\mu}$)
\item $h_{2,1}$ vector multiplets ($A_{\mu}^{I}$; $G^I$)
\item $h_{1,1}$ hypermultiplets ($\phi_A, b_A, c_A, \gamma_A$)
\item the IIB universal hypermultiplet ($\Phi, a , \psi, \chi$)
\ei As shown in \cite{deWit:1984px}, supersymmetry does not allow
couplings between vector and hypermultiplets in the leading $4d$
effective action, if the hypermultiplets are neutral. Since $g_s =
e^{\Phi}$ is in a neutral hypermultiplet, supersymmetry then tells
us that there are no string loop corrections to the tree-level
results. In particular, the metric on the moduli space $\M_V$ of the
vector multiplets (the complex structure moduli for type IIB
compactifications) is exact at tree level, and is read off from the
kinetic terms in the $\N =2$ Lagrangian.

\subsection{IIA Calabi-Yau compactification}
\medskip

We now  repeat the analysis of the previous section for type IIA on
a Calabi-Yau $M$. The $10d$ field content is now given by the
graviton $G_{MN}$, the NS-NS two form $B^{(2)}$, the dilaton $\Phi$,
 an RR two-form field strength $F^{(2)} = d C^{(1)}$ and a four-form $F^{(4)} = d
C^{(3)}$. The dimensional reduction of the metric gives us, as
usual, the $4d$ graviton, $h_{1,1}$ real K\"{a}hler moduli $\phi_A$
and $h_{2,1}$ (complex) complex structure moduli $G^I$. The
reduction of the $B$-field gives $h_{1,1}+1$ scalars $b_A$ and
$\psi$. $C^{(1)}$ will just give one vector field and $C^{(3)}$ will
give rise to $h_{1,1}$ vectors $C^{A}_{\mu}$ and $2 h_{2,1} +2$ real
scalars via the decomposition $C^{(3)} = (\varphi_1+i\varphi_2)
\Omega +(\varphi_1-i\varphi_2) \bar{\Omega} + \varphi_1^I\a_I +
\varphi_2^I \b_I$. One linear combination of the vector fields will
be again the graviphoton. The multiplets we obtain are then \bi
\item the gravity multiplet ($g_{\mu\nu}$, $A_{\mu}$)
\item $h_{1,1}$ vector multiplets ($C^A_{\mu}$, $b^A$, $\phi^A$)
\item $h_{2,1}$ hypermultiplets ($G^I$, $\varphi_{1,2}^I$)
\item  IIA universal hypermultiplet ($\Phi,\psi,\varphi_{1,2}$)
\ei While $h_{1,1} \geq 1$ (the Calabi-Yau must have a (1,1)
K\"{a}hler form), $h_{2,1}$ can be zero (that is a Calabi-Yau with
no complex structure deformations). In that case there is only one
hypermultiplet - the universal one. Note that the numbers of vector
and hypermultiplets we get in type IIA/IIB compactifications are
consistent with mirror symmetry, which exchanges complex structure
and K\"{a}hler moduli.

The dilaton is again in a hypermultiplet, so once more the moduli
space of the vector multiplets does not get corrected by string
loops\footnote{Since in type IIA compactifications the vector
multiplet scalars correspond to K\"{a}hler moduli, the metric on the
moduli space can receive worldsheet $\alpha'$ corrections from
string instantons wrapping the two-cycles in the Calabi-Yau (see
also section 3.2). This is to be contrasted with type IIB, where the
tree-level metric is exact.}.

\subsection{$\mathcal{N}=2$ $4d$ supergravity and special geometry }

\medskip

$\N =2$ supersymmetry highly constrains the form of the Lagrangian.
As far as the scalars are concerned, we already mentioned that the
moduli space of the theory factorizes into a target space $\M_V$
parameterized by the vevs of the vector moduli, and $\M_H$,
parameterized by the hypermultiplet scalar vevs. Supersymmetry
requires that $\M_V$ be a special K\"{a}hler manifold
\cite{deWit:1984px,Strominger:1990pd,Mohaupt:2006ph}, while $\M_H$
is restricted to be a quaternionic K\"{a}hler manifold
\cite{Bagger:1983tt}. In these lectures we will only be concerned
with the action for the vector multiplets. The kinetic terms are determined entirely from the holomorphic prepotential
$F$ of the $\N =2$ theory, which in our context is determined by the
Calabi-Yau geometry and computed from tree level string theory. A
more detailed recent review can be found in \cite{Mohaupt:2000mj},
whose conventions we follow.

\medskip

 Let us now collect a few soon-to-be-needed facts about the leading terms in the
  $\N=2$
 Lagrangian\cite{deWit:1984px,Mohaupt:2000mj,Ceresole:1995ca}, which correspond to
 considering just the tree level term in the Calabi-Yau prepotential ($F=F_0$ in (\ref{fg}); see de
 Wit's lectures for more details). To
 be specific, we assume here that our $\N=2$
 action was
 obtained by compactification of type IIB. In the basis introduced in the previous section,
 the holomorphic three-form can be written as
\be \Omega = X^{\L} \a_{\L} - F_{\L} \b^{\L}. \ee $X^{\L}$ (or
$F_{\L}$) turn out to be projective coordinates on the vector
multiplet moduli space (parameterized by $z^i$, $i \in \{1, \ldots
N_V = h_{2,1}\}$), whose geometry is completely determined by the
choice of holomorphic three-form $\Omega(z^i)$. The K\"{a}hler
potential on $\M_V$ is given by \be \mathcal{K}(z^i,\bar{z}^i)= -
\ln  i \int_M \Omega \wedge \bar{\Omega}, \ee which can be rewritten
as \be e^{-\mathcal{K}} = i (\bar{X}^{\L}F_{\L} - X^{\L}
\bar{F}_{\L}). \ee The periods $F_{\L}$ can be obtained from the
prepotential as \be F_{\L} = \frac{\p F(X)}{\p X^{\L}}. \ee One
choice of (non-projective) holomorphic coordinates on $\M_V$ are
{\it special} coordinates \be Z^I = \frac{X^I}{X^0}\, , \;\;\; I \in
\{1,\ldots h_{2,1} \}. \ee While using these coordinates is useful
for a large number of purposes, please note that the symplectic
invariance of the action is no longer manifest, which may sometimes
not be very convenient. Note also that while $X^{\L}$ and $F_{\L}$
are holomorphic, sometimes (especially in the supergravity
literature) people find it useful to define the rescaled  periods
$(X^{'\L},F'_{\L}) = e^{\mathcal{K} / 2} (X^{\L},F_\L)$, which are
no longer holomorphic.

Another piece of information we need from the $\N=2$ Lagrangian are
the gauge kinetic terms. They read \be \mathcal{L}_{gauge} =
\frac{i}{4} \bar{\N}_{\L\S} \hat{F}_{\mu\nu}^{-\L} \hat{F}^{-\S
\mu\nu} - \frac{i}{4} \N_{\L \S}\hat{F}_{\mu\nu}^{+\L} \hat{F}^{+\S
\mu\nu}, \ee where $\hat{F}^{\pm} =\half( \hat{F} \mp i \star
\hat{F}) \label{gaugekin}$ are the the self-dual and anti-self-dual
parts of the gauge fields\footnote{ In four Lorentzian dimensions,
(anti)self-duality reads $\star \hat{F}^{\pm}= \pm i \hat{F}^{\pm}$.
}. The expression for $\N_{\L\Sigma}$ in terms of the prepotential
and its derivatives is \be \N_{\L\Sigma} = \bar{F}_{\L\Sigma} + i
\frac{N_{\L\Delta} X^{\Delta} N_{\Sigma\Omega}
X^{\Omega}}{N_{\Delta\Xi} X^{\Delta} X^{\Xi}}, \ee where $F_{\L\S}$
denote the second derivatives of the prepotential with respect to
$X^{\L}$ and $X^{\S}$ and $N_{\L\S} = -i(F_{\L\S} -\bar{F}_{\L\S})$.
One can easily show that
 \be F_{\L}(X) = \N_{\L \Sigma} X^{\Sigma} \ee
a relation that will be useful later.

The integral of $\hat{F}$ over a sphere at infinity gives the
magnetic charge of the field configuration we are studying. \be
p^{\L} = \frac{1}{4\pi} \int_{S^2} \hat{F}^\L =  \frac{1}{2\pi} Re
\int_{S^2} \hat{F}^{+\L}. \ee The total electric charge is usually
given by the integral of the dual field strength over the sphere at
infinity, which follows from the Maxwell equation   $ d \star F =0$.
Note that our action (\ref{gaugekin}) gives rise to different
equations of motion, which imply that the electric charge is given
by
 \be q_{\L} = \frac{1}{2\pi} Re \int_{S^2} G_{\L}^+ \ee
where \be G_{\L}^+ = \N_{\L \Sigma} \hat{F}^{ \Sigma+} \ee or
equivalently \be G_{\L} = (Re \N_{\L\S} ) \hat{F}^{\S} + (Im
\N_{\L\S}) \star \hat{F}^{\S} \ee One last thing we would like to
explain is how to recognise which linear combination of the gauge
fields is the graviphoton. We know that the particular combination
has to be symplectically invariant, since which field is in the
gravity multiplet should not depend on the choice of symplectic
basis.

From the field strengths $\hat{F}^{\L}, G_{\L}$ and the periods one
can construct a naturally symplectically-invariant field strength,
given by \be T_{\mu\nu}^- = F_{\L}  \hat{F}^{-\L}_{\mu\nu} - X^{\L}
G_{\L}^- ,\ee which corresponds to the graviphoton. The graviphoton
is also special in that its charge \be Q = \frac{1}{4\pi} \oint T^-
\ee is proportional to the central charge of the $\N=2$
supersymmetry algebra\footnote{Note that the Q so defined rescales
under projective transformations.}. If we restrict our attention to
just supersymmetric solutions (preserving 4 or 8 supercharges), then
\be F_{\L} \hat{F}^{+\L} - X^{\L} G_{\L}^+ =0 \ee when evaluated on
these solutions, so the graviphoton charge becomes \be Q = F_{\L}
p^{\L} - X^{\L} q_{\L} = Q_{mag} + i Q_{el}, \ee which generally is
complex.

Having talked at length about electric and magnetic charges in the
$4d$ theory, we might as well mention how to obtain charged
particles in string theory compactified on a Calabi-Yau $M$. The
answer is simple: just wrap D-branes on the various cycles in $M$.
The string theory D-branes source the $10d$ RR fields, which from the $4d$
point of view look like pointlike charges that source the
different gauge fields which come from the dimensional reduction of
the $10 d$ RR fields. In type IIB, D3-branes can wrap any of the $2
h_{2,1} + 2$ different 3-cycles, giving a total of $2 h_{2,1} + 2$
different electric and magnetic charges in $4d$. The number of units
of $4d$ charge is determined by how many times we wrap the
$D$-branes around the particular cycle.

In type IIA the stable D-branes are even-dimensional. One can again
produce any electric and magnetic charges by wrapping D6, D4, D2 and
D0 branes on the various cycles in the Calabi-Yau. If $A \in \{1,
\ldots, h_{1,1}\}$ labels the 2 (and also the dual 4)-cycles, then
the most general set of charges we can get is $(p^0, p^A, q_A,
q_0)$, which stands for D6, D4 (magnetic) and D2, D0 (electric)
charges respectively.

If we wrap a  large number of branes at the same point in noncompact
$4d$ space, we will have to consider the backreaction of the metric
and the other supergravity fields. It turns out that for large
charges one can obtain macroscopic black holes
\cite{Mohaupt:2000gc,Peet:2000hn}, which we will now turn to study.

\subsection{Black hole solutions}

\medskip

In a classic paper \cite{Gibbons:1982fy}, Gibbons and Hull have
shown that minimal $4d$ $\N=2$ supergravity (whose bosonic sector is
just Einstein - Maxwell gravity) has charged black hole solutions of
the Reissner-Nordstr\o m type. These solutions are parametrised by
their mass $M$ and charge $Q = Q_{mag} + i Q_{el}$. Since $Q$
happens to equal the central charge of the $\N=2$ supersymmetric
theory, then the BPS bound requires that these black holes have $M
\geq |Q|$. If $M>|Q|$ their Hawking temperature is nonzero, which
means
 they can radiate and thus are not stable objects. Since we are
looking for supersymmetric solutions, which have to be stable, the
remaining candidate is the extremal solution with $M=|Q|$. The
Hawking temperature of these black holes is zero and the solution
indeed turns out to be supersymmetric. These extremal objects have
been the focus of many interesting investigations over the years.

The easiest way to see whether a particular solution is
supersymmetric is to look at the fermion variations for that
particular background and require that they vanish. The only
fermions present in minimal supergravity are the gravitini, whose
variation is \be \delta_{\e} \psi_{\mu}^{\a} = 2 \nabla_{\mu}
\e^{\a} - \frac{i}{8} T_{\lambda \nu}^- \gamma^{\lambda \nu}
\gamma_{\mu} \e^{\a\b} \e_{\b} .\label{dpsimin}\ee The solution will
be supersymmetric if there exists a spinor $\e^{\a}$ such that
$\delta_{\e} \psi_{\mu} =0$. It turns out that there exist four such
spinors, so the solution preserves half of the original eight
supersymmetries.
\begin{figure}[htp]
\centering
\includegraphics[height=6 cm]{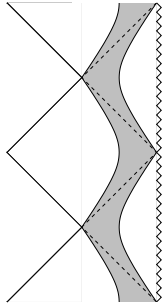}
\caption{Penrose diagram for the extremal Reissner-Nordstr\o m black
hole. The zigzag stands for the timelike singularity (at $r=-Q$) and
the dotted lines represent event horizons. The shaded region covers
the near-horizon $AdS_2 \times S^2$, illustrated in figure
3.}\label{rnbh}
\end{figure}

The expression for the metric is \be ds^2 =  - e^{2U(r)} dt^2 +
e^{-2U(r)} (dr^2 + r^2 d\Omega^2_2), \label{metric} \ee where \be
e^{-U(r)} = 1 + \frac{|Q|}{r}. \ee The solution carries charge $Q$, as
the expression we get for the graviphoton indicates  \be
T_{\mu\nu}^- = Q \,\e_{\mu\nu}^- \;\;\;,\mbox{with}\;\;\;\;
\int_{S^2} \e^- = 4 \pi .\ee As $r \r \infty$, $e^U \r 1$ and the
metric becomes just the flat metric on $\mathbb{R}^{3,1}$. As $r \r
0$ the metric takes the form \be ds^2 = - \frac{r^2}{|Q|^2} dt^2 +
\frac{|Q|^2}{r^2} dr^2 + |Q|^2 d\Omega_2^2. \ee The near-horizon
geometry is thus $AdS_2 \times S^2$, and the area of the horizon is
$\mathcal{A} = 4 \pi |Q|^2$. The near horizon bosonic isometry group
is $SO(2,1)\times SU(2)$, which is part of the $SU(1,1|2)$
superisometry group containing the maximal eight supersymmetries.
This means that $AdS_2 \times S^2$ is a maximal $\N=2$ vacuum, and
hence the Reissner-Nordstr\o m solution we obtained can be thought
of as a soliton radially interpolating between two maximally
supersymmetric vacua \cite{Gibbons:1993sv}.

\begin{figure}[htp]
\centering
\includegraphics[height=4 cm]{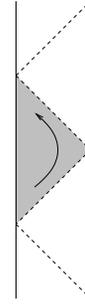}
\caption{Penrose diagram for $AdS_2$. The dotted lines represent the
horizons inherited from the embedding in the extremal
Reissner-Nordstr\o m geometry of figure 2. }\label{ads2}
\end{figure}

\subsection{The attractor equations}
\medskip

Next, we add in vector multiplets
\cite{Ferrara:1995ih,Strominger:1996kf}. The gravitini variations
now acquire extra terms, which contain derivatives of the vector moduli

 \be \d_{\e}
\psi_{\mu}^{\a} = 2 \nabla_{\mu} \e  - \frac{i}{8} T^-_{\nu\lambda}
\gamma^{\nu\lambda}\gamma_{\mu} \e^{\a\b} \e_{\b}+ i A_{\mu}
\e^{\a}, \ee  where \be A_{\mu} =  \frac{i}{2} N_{\L\S}
(\bar{X}^{\L} \p_{\mu} X^{\S} - \p_{\mu} \bar{X}^{\L} X^{\S})\ee
where in the above two and the following equations we have fixed the
gauge $N_{\L\S} X^{\L} \bar{X}^{\S}= - 1$.

 The supersymmetry variations of the fermions in the vector multiplets
(the gaugini) are \bea \d \Omega^{\L}_{\a} &=& 2 \gamma^{\mu}
(\p_{\mu} + i A_{\mu} )X^{\L} \e_{\a}  \non \\ &+& \half
\gamma^{\nu\lambda} (\hat{F}^{+ \L}_{\nu\lambda} - \frac{1}{4}
X^{\L} T^{+}_{\nu \lambda})  \e_{\a\b} \e^{\b} \eea We would like to
see if a maximally supersymmetric near-horizon $AdS_2 \times S^2$
region with an $SU(1,1|2)$ superisometry group can still exist. This
requires that the moduli $X^{\L}$ be constant throughout the
near-horizon spacetime. Then the gravitini variations vanish as
before and the gaugino equations require that \be \hat{F}^{+
\L}_{\mu\nu} = \frac{1}{4} X^{\L} T^{+}_{\mu\nu}. \ee Integrating
our previous solution for the graviphoton $T_{\mu\nu}^+ \propto
\e_{\mu\nu}^+$ (where $\e^+$ is a self-dual two-form) over the
horizon $S^2$ we find \be p^{\L} = \frac{1}{2\pi} \int_{S^2} Re \,
\hat{F}^{+\L} = Re[CX^{\L}] \label{patt}, \ee for some complex
constant $C$, which in general will depend on our choice of
projective gauge\footnote{For $N_{\L\S} X^{\L} \bar{X}^{\S}= - 1$,
we have $\bar{C}= 2 i Q$.}. Similarly, \bea q_{\L} &=&
\frac{1}{2\pi} \int_{S^2} Re \, \hat{G}^{+}_{\L} \non =
\frac{1}{2\pi} \int_{S^2} Re \,\N_{\L\Sigma} \hat{F}^{+\Sigma} \\
&=& Re[C \N_{\L\Sigma} X^{\Sigma}] = Re[C F_{\L}]. \label{qatt} \eea
We thus see that the moduli $X^{\L}$, $F_{\L}$ must take very
specific values in terms of the black hole charges in order to
recover the $AdS_2\times S^2$ part of the solution in the
near-horizon region. For reasons which will become clear shortly
these equations are known as the {\it attractor equations}.

The attractor equations, together with knowledge  of the
prepotential $F(X^{\L})$, generically let
us determine in principle (up to possible discrete choices) the real
and imaginary parts of all $CX^{\L}|_{hor}$ in terms of the
charges of the black hole. This can be easily checked by counting the
number of equations and unknowns.

The Bekenstein-Hawking entropy depends only on the near-horizon data
and is simply given by the horizon value of \be S_{BH} = \frac{\pi
i}{2} (q_{\L} \bar{C} \bar{X}^{\L} - p^{\L} \bar{C} \bar{F}_{\L}).
\label{sbh} \ee After solving the attractor equations, the entropy
becomes just a function of the charges of the black hole, and does
not depend on any asymptotic data, such as the asymptotic values of
the moduli (as long as we do not leave the basin of attraction).

In general one might want to find the full black hole solution,
including the asymptotic region. This is a harder problem, and one
we will not need to solve for now because the full solution has the
form of a radially interpolating soliton, with the moduli in the
maximally-symmetric near horizon region determined by the charges of
the black hole via the attractor equations. So, given the charges
and the value $X_{\infty}^{\L}$ of the moduli at infinity, the
variation through space of $X^{\L}$ will be such that it always goes
to the horizon value $X^{\L}_{hor}$ determined by the attractor
equations \cite{Ferrara:1996um}. The path traced in moduli space by
$X^{\L}(r)$ \footnote{ We here only consider spherically-symmetric
solutions. It turns out there also exist multicenter solutions
\cite{Bates:2003vx}.} is called an {\it attractor flow}.

\begin{figure}[htp]
\centering
\includegraphics{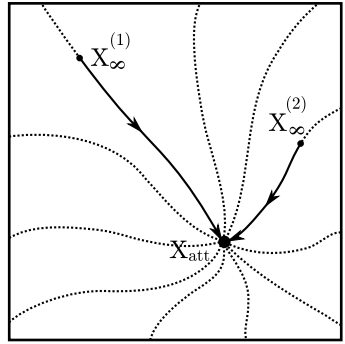}
\caption{Attractor flow in moduli space. No matter which point
(within the basin of attraction) in the moduli space at $r = \infty$
we choose as our asymptotic data, at the horizon ($r=0$) the moduli
always take the same attractor value, determined by the charges of
the black hole.}\label{attflow}
\end{figure}

While at first this behavior of the vector moduli might seem a bit
surprising, it is actually required for the existence of sensible
counting of microstates of the black hole: the area of the horizon
depends on $X^{\L}_{hor}$, but at the same time it is the logarithm
of the number of microstates of the black hole - that is - the log
of an integer. Therefore one does not expect it to smoothly depend
on continuous parameters such as the asymptotic values of the
moduli.\footnote{However a more refined analysis shows that it can
depend $discontinuously$ on the asymptotic
moduli\cite{Bates:2003vx}. This leads to very interesting
modifications of the simplified picture presented here.
 See
\cite{Denef:2007vg} for a recent discussion.} Additionally, the
entropy is not expected to depend on parameters that we may choose
to tune at infinity, since it is an intrinsic property of the black
hole. Note that there is no attractor mechanism for the
hypermultiplet moduli, but this does not affect the above reasoning,
since the horizon hyperscalar vevs drop out of the entropy
formula\footnote{In recent work by Sen \cite{Sen:2005wa}, it was
shown that all fields that the entropy depends on at the horizon
must be subject to an attractor mechanism. His reasoning works not
only for supersymmetric black holes, but any black hole whose
near-horizon symmetry group contains $SO(2,1)$
\cite{Astefanesei:2006dd,Dabholkar:2006tb,Kallosh:2005ax,Kallosh:2006bt,Kallosh:2006ib,Sahoo:2006rp}.}.

\subsubsection{  A simple example}
\medskip

Maybe the simplest example \cite{Shmakova:1996nz} is to consider IIA
compactified on a Calabi-Yau with moduli $X^{\L}= (X^0,X^A)$. In the
large-volume limit we can take the Calabi-Yau prepotential to be \be
F(X) = D_{ABC} \frac{X^A X^B X^C}{X^0} ,\label{treep}\ee where $6
D_{ABC}$ are the intersection numbers of the Calabi-Yau. In that
case we have \bea
F_A &=& \p_A F = \frac{3 D_{ABC} X^B X^C}{X^0} \non \\
F_0 &=& \p_0 F = - \frac{D_{ABC}X^A X^B X^C}{(X^0)^2.} \eea We will
consider a $D4$-$D0$ black hole, with charges $p^0=q_A =0$ and $p^A,
q_0 \neq 0$. The attractor equations read \bea
&& Re[CX^A] = p^A, \;\; \; Re[CX^0] = 0, \non \\
&& Re[CF_A] = 0 \;, \;\;\;\;\; Re[CF_0] = q_0. \eea It is easy to
see that the solution to the above equations is \be CX^A = p^A \;,
\;\;\; CX^0 =  i \sqrt{\frac{D}{q_0}} ,\ee where we have defined $D
\equiv D_{ABC} p^A p^B p^C$. From (\ref{sbh}) we can compute the
macroscopic entropy associated with this black hole
\be S_{BH} = \frac{\pi i}{2} (q_0 \bar{C} \bar{X}^{0} - p^A \bar{C}
\bar{F}_{A}) = 2 \pi \sqrt{D q_0}. \label{d4d0e} \ee

\medskip

\subsection{The general asymptotically flat solution}

Somewhat surprisingly, given a solution of the near-horizon
attractor equations for the vector moduli as a function of the
charges, it is possible to actually construct the full
asymptotically flat solution with generic values for the moduli at
infinity. Here we will concentrate only on the spherically symmetric
case, although generic exact multi-center solutions are also known
\cite{Bates:2003vx}.

Due to spherical symmetry, the metric must take the form
(\ref{metric}). The insight of
\cite{Behrndt:1997ny,Sabra:1997dh,Sabra:1997kq}, was to notice that
the moduli fields that satisfy the BPS equations of motion
\cite{Ferrara:1997tw} must obey equations that are very similar to
the attractor equations (\ref{patt}) and (\ref{qatt}), just that
they hold throughout spacetime, and not only at the horizon of the black
hole
 \bea & Re [C(r) X^{\L}(r)] = H^{\L}(r) \non \\ &
 Re[C(r)F_{\L}(X(r)]=
H_{\L}(r) .\eea Here $H^{\L}, H_{\L}$ are harmonic functions on flat
$\mathbb{R}^3$ sourced by the charges $p^{\L}, q_{\L}$. It follows
that whenever one can solve the attractor equations (which is
generally a difficult task), one can also get the solution for the
moduli everywhere\footnote{When the attractor equations are not
analytically soluble, spacetime solutions can still be explicitly
written in terms of the implicit attractor solutions.}. This is done
with the aid of the {\it entropy function}, $\S$, defined as \be
 \S (\vec{x}) = \frac{1}{2} Im [CX^{\L} (\vec{x}) \bar{C}
 \bar{F}_{\L}(\vec{x})].
\ee By comparison with (\ref{sbh}), we see that near the horizon $
\S(r, p^{\L}, q_{\L}) = (\pi r^2)^{-1} S_{BH}(p^{\L}, q_{\L})$. Away
from the horizon, $\S(r)$ is simply given by making the replacement
\bea
p^{\L} &\r& H^{\L}(r) = \frac{p^{\L}}{r} + h^{\L} \non \\
q_{\L} &\r&  H_{\L}(r) = \frac{q_{\L}}{r} + h_{\L} \eea in the
entropy formula $\frac{1}{\pi} S_{BH} (p^{\L}, q_{\L})$.  The
constants $h^{\L}, h_{\L}$ can be determined from the asymptotic
values of the moduli. Finally, the solution for the metric and the
moduli is \be e^{-2U} = \Sigma(H) \;, \;\;\;\; C X^{\L}= H^{\L} + i
\frac{ \p \Sigma(H)}{\p H_{\L}}. \ee This is a very powerful result,
since it allows us to reconstruct the solution for the metric and
moduli {\it throughout spacetime} from just knowledge of the entropy
as a function of the charges and the asymptotic values of the
moduli.

There is a catch though, in that even if the entropy itself is real,
the entropy function is not guaranteed to be so. $\S^2(r)$ is a
quartic polynomial in $\frac{1}{r}$, which is positive as $r\r
0,\infty$, but there is nothing to prevent it from becoming negative
for an intermediate range of $r$, if we tune the moduli at infinity
appropriately. Since $\S(r)$ is a metric component, the solution
becomes unphysical if it becomes imaginary. Therefore one must
always check whether our formal solution is actually physical by
making sure that $\S(r)$ is real and the moduli $\tau^A(r)$ belong
to the physical moduli space for all values of the radius.

Even if the asymptotic moduli are outside the regions for which the single-centred solution exists, the respective BPS
state might still be realized as a black hole (or point particle)
bound state \cite{Denef:2000nb}. The entropy function can be adapted \cite{Bates:2003vx} to describe these
 multicenter supersymmetric  black holes. One
simply allows the harmonic functions to have poles at the location
of each black hole carrying the corresponding charge. These
solutions have many interesting properties and applications
\cite{Gaiotto:2005gf,Denef:2000nb,Denef:2001xn,Gaiotto:2005xt,Bena:2005ni,Dijkgraaf:2005bp,Denef:2007vg},
but are outside the scope of these lectures.

Note that checking the reality of $\S(\vec{x})$ in the case of
multicenter solutions is an extremely difficult task even for simple
Calabi-Yaus, which could only be tackled numerically.

\subsubsection{Simple example redux}
\medskip

 As an example, let us again take the D4-D0 black
 hole, but now in a Calabi-Yau compactification that has only one complex K\"{a}hler modulus $\tau(\vec{x})$, where $\vec{x}$ denotes position in $\mathbb{R}^3$, and the triple self-intersection number of the Calabi-Yau is $D_{111}=1$.
 The entropy is then $S = 2 \pi \sqrt{q_0 p^3}$. The entropy function is
\be \S(H(\vec{x}), H_0(\vec{x})) = 2 \sqrt{H_0(\vec{x})
H^3(\vec{x})} = e^{-2 U(\vec{x})} ,\ee where the harmonic functions
are \be H= \frac{p}{r} + h \; , \;\;\;  H_0 = \frac{q_0}{r} + h_0
.\ee The solution for the modulus in this case is purely imaginary
\be \tau(\vec{x}) =\frac{CX(\vec{x})}{CX^{0}(\vec{x})}= -i
\sqrt{\frac{H_0(\vec{x})}{H(\vec{x})}} \ee so now we only need to
determine the constants $h_0, h$ in terms of the asymptotic value of
the modulus $\tau_{\infty} = - i a_{\infty}$ (here the physical
moduli space is the lower half plane). We choose a gauge so that as
$r \r \infty$ the scale factor $e^{-2U} \r 1$, which imposes the
constraint $2 \sqrt{h_0 h^3} = 1$. We then get $h_0 = 2^{-\half}
(i\tau_{\infty})^{3 \over 2}$, $h= ( 2 i\tau_{\infty})^{-\half}$.
Note that the asymptotic K\"{a}hler class $i\tau_{\infty}$ has to be
positive for the solution to exist.

The attractor equations only require that the B-field
(proportional to $Re \tau$) at the horizon be zero, but its value at infinity does not have to vanish,
as our solution seems to indicate. Nonzero $B_{\infty}$ can be obtained by considering the entropy
function associated to a black hole with additional $D2$ charges, but in which we take the
corresponding $H_{A} = constant$.

\subsubsection{Generic entropy for cubic prepotential}

We now give the formula \cite{Shmakova:1996nz} for the entropy of
the generic D6-D4-D2-D0 black hole when the prepotential is simply
\be F=\frac{D_{ABC}X^AX^BX^C}{X^0} ,\ee  Then the entropy of the
black hole with charges $(p^0, p^A, q_A, q_0)$ is \be S = 2 \pi
\sqrt{Q^3 p^0 -  J^2 (p^0)^2} \label{entropy} \ee where $Q$ is
determined by solving the following equations for a set of variables
$y^A$ \be 3 D_{ABC} y^A y^B = q_A + \frac{3 D_{ABC} p^B p^C}{p^0}
\label{eqya}\ee which for general charges and intersection numbers
are not analytically soluble. In any case, we have \be
 Q^{3\over 2} = D_{ABC} y^A y^B y^C
\ee and \be J = - \frac{q_0}{2} +  \frac{D_{ABC} p^A p^B
p^C}{(p^0)^2} + \frac{p^A q_A}{2 p^0} .\ee

 Please note that our analysis is only valid when the Calabi-Yau is large both at
 infinity and at the horizon, since the entropy formula was derived using the
 large volume
 prepotential (\ref{treep}). One has to pay special attention if the attractor flow in question passes
 through a region in moduli space where the Calabi-Yau (or some cycle in the Calabi-Yau) is becoming
 small, since then instanton corrections to the prepotential become important \cite{Denef:2007vg}.
 In order for our supergravity analysis to be valid, we also need the $4d$ curvature to be small
 everywhere, which translates into the requirement of large black hole charges.

\bigskip
\subsection{Higher orders }
\medskip
\subsubsection{Corrections to the $\N =2$ action}
\medskip
In this subsection we briefly sketch how the $\N=2$ invariant
Lagrangian, including higher curvature corrections, is constructed
using the so-called superconformal tensor calculus
\cite{deWit:1984px}. The idea is to begin by constructing an action
for $\N=2$ conformal supergravity coupled to $N_V+1$ conformal
vector multiplets and then gauge-fix it down to Poincar\'{e}
supergravity coupled to $N_V$ vector multiplets. The advantage of
this approach relies on the fact that more symmetries are realized
in a linear and simple way in conformal supergravity, and also the
off-shell multiplets are smaller\footnote{Indeed the simplifications
are so striking one suspects there may be some deeper physical
significance underlying this "mathematical trick".}. The basic
ingredients are two types of superconformal multiplets \bi
\item the Weyl multiplet, which contains the vierbein $e_{\mu}^a$ and the
two gravitini, among many other auxiliary fields. To incorporate it
into the action, one actually has to construct an $\N=2$ chiral
multiplet $W^2$, which contains the gauge-invariant field
strengths\footnote{Very roughly, the highest component of $W^2$ is
the antiselfdual part of the Weyl tensor squared, while the lowest
component is the square of the antiselfdual part $(T^-)^2$ of an
auxiliary tensor field $T$ that gets identified with the graviphoton
upon gauge-fixing.}
\item $N_V+1$ vector multiplets, each containing a scalar $X^{\L}$, a gaugino $\Omega^{\L}$ and a gauge field $A_{\mu}^{\L}$, among other stuff. One of these vector multiplets will provide the graviphoton for the Poincar\'{e} gravity multiplet, since the Weyl multiplet does not contain an independent gauge field.
\ei The action for the vector multiplets ${\bf{X}}^{\L}$ (where
boldface type denotes a superfield) is constructed by taking a
holomorphic function $F({\bf{X}}^{\L})$ and integrating it over
$\N=2$ superspace. Dilation invariance requires that $F$ be
homogenous of degree two in the ${\bf{X}}^{\L}$. To recover minimal
supergravity we should only consider one vector multiplet $\bf{X}$,
for which the prepotential is $F(\bf{X}) ={\bf X}^2$. Integrating
over superspace we obtain \be \int d^4 \theta \, F({\bf X}) = X
(\Box_4 - \frac{1}{6} R) \bar{X} + \ldots .\ee Upon conformal
gauge-fixing $X$=const this gives rise to the Einstein-Hilbert term.
The Maxwell term comes about in the usual way
\cite{Alvarez-Gaume:1996mv}. A similar mechanism will give rise to
the Einstein-Hilbert, Maxwell and scalar terms in the non-minimal
action. Note that we will get only $N_V$ scalars out of the initial
$N_V+1$, since one combination gets gauge-fixed to a constant.

In order to include higher curvature terms in the Lagrangian, one
has to add couplings to the chiral multiplet $W^2$. This is achieved
by simply extending the holomorphic prepotential to also be a
function of $W^2$, $F({\bf{X}}^{\L}, W^2)$, which can be expanded as
\be\label{fg} F({\bf X}^{\L}, W^2) = \sum_{g=0}^{\infty} F_{g} ({\bf
X}^{\L}) \, W^{2g}\ee The $F_g$ are now required to be homogenous of
degree $2-2g$. Quite interestingly, they are related to topological
string genus $g$ amplitudes, as we will explore later in these
notes.

\subsubsection{Wald's formula}
\medskip
The Bekenstein-Hawking area law for the macroscopic entropy of a
black hole  was derived in Einstein gravity. It cannot possibly
remain exactly valid when higher
 curvature
 corrections to the action are included,  as the area is not invariant under field redefinitions (e.g. which mix
 $g_{\mu\nu}$ and $R_{\mu\nu}$), while the entropy must be. It was
shown in \cite{Wald:1993nt,Jacobson:1993vj} how the area law has to
be modified in the presence of $R^2$ or higher derivative terms in
the Einstein action in order that the first law of black hole
mechanics - the spacetime manifestation of the first law of
thermodynamics - remain valid. The first law of black hole mechanics
can be put in the usual form \be \d M = \frac{\kappa_S}{2\pi}  \d S
+ \phi \d q + \mu \d p ,\ee where $\kappa_S$ is the surface gravity
on the horizon, if $S$ is given by\footnote{This is the correct
formula in the case in which the effective action contains powers of
the Riemann tensor, but no derivatives thereof.} \be S = 2 \pi
\int_{\mathcal{H}}  \e_{\mu\nu} \e_{\rho\sigma} \frac{\p
\mathcal{L}}{\p R_{\mu\nu\rho\sigma}} \, d\Omega .\label{waldf}\ee
Here $\e_{\mu\nu}$ is the binormal to the horizon $\mathcal{H}$ and
$d \Omega$ is the volume element on $\mathcal{H}$. Note that if the
Lagrangian only consists of the Einstein-Hilbert term, then $S$
equals the area of the horizon, but $R^2$ and higher curvature
corrections to the action do generically modify the area law.

 Thus, upon adding higher curvature terms to the $\N=2$ supergravity action,
 the entropy of the black hole solutions gets modified in two ways: first, the metric on
 the horizon changes as a consequence of the modified equations of motion. Second, the entropy
 formula itself receives corrections according to (\ref{waldf}).

 The case that concerns us - of BPS black holes in $\N=2$ supergravity
 coupled to vector multiplets - was considered
 in \cite{Behrndt:1996jn,LopesCardoso:1998wt,LopesCardoso:2000qm,LopesCardoso:1999cv,LopesCardoso:1999xn}. The authors
 argued that after adding the terms (\ref{fg}) to the $\N=2$ supergravity action, the near horizon geometry was still $AdS_2 \times
S^2$, and that the moduli were still subject to an attractor
mechanism. Their horizon values were fixed by the following
generalisation of the attractor equations \be Re \,CX^{\L} = p^{\L},
\;\;\; Re \,C F_{\L} = q_{\L} \label{genatt} \ee \be C^2 W^2 = 256
\label{watt} \ee where now $F_{\L}$ is the derivative of the full
corrected prepotential $F(X^{\L},W^2)$. Taking into account the combined effect of the metric backreaction and
 Wald's corrections to the area
law, the expression for the entropy becomes \be S_{BH} =
\frac{\pi}{2} Im (C X^{\L} \bar{C} \bar{F}_{\L}) - \frac{\pi}{2} Im
(C^2 W \p_{W} F) \ee evaluated on the horizon. In lecture 3 we will
use this result to match the perturbative expansion of the black
hole partition function to the perturbative expansion of the
topological string.

\subsection{The MSW CFT}

\medskip

In this subsection we discuss the CFT duals known for a large class
of Calabi-Yau black holes. Finding the dual for the most general
case is a very interesting unsolved problem. A dual
 CFT description \cite{Maldacena:1997de} (often known as the MSW CFT) has been proposed in the still fairly general case
 in which the D6 charge is zero, but the D4-D2-D0 charges and the Calabi-Yau itself
 are (almost)
 arbitrary. We will denote the charges by $(p^A, q_A, q_0)$ and the intersection numbers of the Calabi-Yau $M$
 by $ 6 D_{ABC}$. As before, we define $D = D_{ABC} p^A p^B p^C$.

 Upon lifting to M-theory, the D4-D2-D0 brane configuration
 becomes an M5-brane wrapped on $P\times S^1$, where the surface $P$\footnote{
 The charges must be such that $P$ is very ample
 \cite{Maldacena:1997de},
 which is the case if the K\"{a}hler class is large.} is holomorphically embedded in $M$ (as a consequence of supersymmetry)
  and $S^1$ stands for the M-theory circle all throughout this lecture. $P$ decomposes as $P = p^A \Sigma_A$ where $\Sigma_A$ is an integral basis of
  4-cycles on $M$. The M5 carries worldvolume fluxes that give rise to
  induced M2-brane charges ($q_A$), as well as $q_0$ units of momentum along the $S^1$.
  From the supergravity point of
  view, under the M-theory lift the D4-D2-D0 black hole becomes an $S^1$-wrapped
  black string, whose near
  horizon geometry is locally $AdS_3 \times S^2\times M$. Note that the
  M-theory lift of a D6-brane is Taub-Nut space,
  so if we considered black holes with D6 charge the eleven-dimensional geometry would no longer be $AdS_3$,
  which is partly the reason that a microscopic description is not known in that case\footnote{Quite surprisingly,
  the MSW string remakes its appearance for certain black holes with nonzero D6 charge \cite{Guica:2007gm}, which may
  imply that {\it all} IIA black holes are described by some deformation of the MSW CFT. This makes it all the
  more worth studying, of course.}.

  The low energy dynamics on the M5 worldvolume is captured by an effective $2d$ CFT living on
  $S^1 \times \mathbb{R}$ (time). This $2d$ CFT has $(0,4)$ supersymmetry, inherited via dimensional
  reduction from the $(0,2)$ supersymmetry  of the $6d$ theory living on the M5 worldvolume.
  Its low-energy excitations arise as zero-modes of the fluctuations of the M5 worldvolume
  fields (embedding, self-dual 3-form $h^{(3)}$ and the right-moving (RM)  $6d$ fermions $\psi^{(6)}$) on $P$, as follows
\bi
\item zero-mode fluctuations of the embedding correspond to cohomology classes on $P$.
\item zero-modes of the $h$-field correspond to self-and anti-self-dual forms on $P$, as can be seen from the decomposition
$$ h^{(3)} = d \phi^A \wedge \a_A \;, \;\;\; \a_A \in H^2(P,\mathbb{Z}) $$
If the 2-form $\a_A$ is self-dual, then the self-duality of $h$
implies that the scalar modulus $\phi^A$ has to be right-moving,
while if $\a^A$ is anti-self-dual, then $\phi^A$ is left-moving.
Thus we obtain $b_2^-$ LM and $b_2^+$ RM scalars in the CFT from the
dimensional reduction of the $h$-field, where $b_2^{\pm}$ are the
numbers of self-dual and respectively anti-self-dual two-forms on $P$.
\item  fermions in the CFT arise from $(0,2)$ forms on $P$ and they are RM. This can be easily seen
by decomposing the $6d$ RM fermions as
$$ \psi_{(6)} = \sum_I \psi_{(2)}^I \otimes \psi_{I}^{P} $$
where $\psi_I^P$ are fermionic zero-modes on $P$, which are known to
be in one-to-one correspondence with harmonic $(0,2)$
forms.\footnote{There are no left-moving fermions because they would
be in one-to-one correspondence with $(0,1)$ forms on $P$, of which
there are none since  $b_1(P)= b_1(M)=0$.} The number of RM fermions
is $4 h_{2,0}(P)$ and it can be shown to equal the number of RM
bosons, as required by supersymmetry on the right.
\item there is one distinguished $\N=4$ multiplet in the $2d$ CFT,
called the {\it centre of mass multiplet}. Its bosonic content is
given by the three massless scalars $X^i$ that parameterize the
motion of the black hole as a whole in the three noncompact
directions and one right-moving mode $\varphi$ of the $h$ field,
which corresponds to the unique self-dual form on $P$ which is
extendible to a 2-form on $M$. This is of course the pullback of the
K\"{a}hler form on $M$ - $J$ - which has to be proportional to $[P]$
at the horizon, as a consequence of the attractor equations. In
terms of the scalars $\phi^A$ \be \varphi = p^A D_{AB} \phi^B \equiv
p^A \phi_A .\ee The fermionic parteners of  these four bosons are
the goldstinoes $\tilde{\psi}^{\pm\pm}$
 that arise from the four supersymmetries
broken by the brane configuration. \ei

 The resulting central charges, including a
 subleading correction proportional to the second Chern class of $M$,
 are \cite{Maldacena:1997de}
\be c_L = 6 D + c_{2}\cdot P\, , \;\;\; c_R = 6 D + \half c_{2}
\cdot P .\ee

The MSW CFT reproduces the area-entropy law. If one is only
interested in the D4-D0 system, then the left-moving oscillator
momentum is $q_0$, while the right-moving oscillator momentum has to
be zero by supersymmetry. Using Cardy's formula, the entropy reads
\be S = 2 \pi \sqrt{\frac{c_L q_0 }{6}} = 2 \pi \sqrt{D q_0} \ee in
agreement with the macroscopic formula (\ref{d4d0e}).

What happens if we add D2/M2 charges? This corresponds to turning on
M2 brane fluxes (that is, nonzero $h^{(3)}$ flux on cycles of the
form $S^1 \times \a$, with $\a$ a two-cycle on $P$) on the M5
worldvolume. In the effective $2d$ theory, membrane charge is the
zero-mode momentum $q_A = \int d\phi_A$ carried by the massless
scalars that arise from the dimensional reduction of the three-form
$h$, and thus it is a vector in the Narain lattice of massless
scalars. $q_A$ contributes to the $S^1$ momentum along the string.
The effect is to shift the momentum available to be distributed
among the LM oscillators by \be
 q_0 \r \hat{q}_0= q_0 + \frac{1}{12} D^{AB} q_A q_B
,\ee $D^{AB}$ is the inverse of the matrix of charges $D_{AB} =
D_{ABC} p^C$. The entropy gets modified to $S= 2 \pi \sqrt{\hat{q}_0
D}$, and it is straightforward to check that this agrees with the
supergravity formula (\ref{entropy}) for $p^0$=0.

\subsection{ The (modified) elliptic genus}

\medskip
So far we have been loosely speaking about the ``entropy'' of the
black hole and its CFT dual. In fact, what we are really looking for
is a BPS protected quantity which does not change as one
extrapolates between the supergravity and CFT regimes. Such an
object is the (modified) elliptic genus, which is an index that
counts a weighted number of ground states. At leading order (given
by the area law), which quantity we use seems not to matter
much\footnote{Presumably what happens is that one or both of $n_B$
and $n_F$ are proportional to $e^{{Area}/{4G}}$, but with different
proportionality constants, so that the difference between the log of
the total number of states and the log of the index is subleading.},
but at subleading orders we will definitely need to be more
precise.\footnote{Actually, the first subleading correction to the
entropy is included in (\ref{entropy}) expression and matches
\cite{Maldacena:1997de}: on the microscopic side it involved the
$c_2 \cdot P$ correction to the central charge, and on the
macroscopic side it came from the Gauss-Bonnet term  in the $4d$
effective action.}

Let us give a very simple example of what an index is
\cite{Witten:1982df}. Take supersymmetric quantum mechanics with one
supercharge $Q$ ($Q^2 = H$), and define the Witten index as \be I_F
= Tr_{states} (-1)^F e^{-\b H} \label{witind} \ee where $F$ is the
fermion number operator. Since for states of nonzero energy $|E
\rangle$, $Q| E \rangle$ is a state with the same energy but
opposite fermion number (mod 2), $I_F$ only gets contributions from
the ground states, since they are annihilated by $Q$ and thus are
not paired. If we denote by $n_B$ and $n_F$ the number of bosonic
and respectively fermionic ground states, then $I_F = n_B - n_F$.
The index is rigid under small deformations of the parameters in the
Hamiltonian.

This simple index actually vanishes for the MSW CFT
\cite{Maldacena:1999bp,Gaiotto:2006wm}, whose symmetry algebra on the right is a
Wigner contraction of the large $\N=4$ superconformal algebra. This
consists of a small $\N=4$ superconformal algebra plus four bosonic
and four fermionic generators, which are nothing but the fields in
the center-of-mass multiplet $(X^i, \varphi, \tp^{\pm\pm})$. This
multiplet has equal numbers of bosonic and fermionic excitations and
hence gives a prefactor of zero for the Witten index.

 The generators of the small $\N=4$ are four supercurrents $\tilde{G}^{\pm\pm}$, three bosonic currents $J^i_R$ that
 generate a level $k$ $SU(2)$ Ka\v{c}-Moody algebra, and the usual Virasoro generators with central charge $c=6k$. We
 have as usual two choices of boundary conditions for the fermionic operators, periodic (R) or antiperiodic (NS).
The  small $\N=4$ generators have certain commutation relations
\cite{Gaiotto:2006wm} with the
 fermionic operators $\tp^{\pm\pm}$ and $\tilde{J}^{\varphi} = \bar{\p} \varphi$.
 We will only need the R-sector commutators
\be \{\tilde{G}^{\a a}_0,\tp_0^{\b b} \} = \e^{\a\b} \e^{ab}
\tilde{J}_0^{\varphi}, \;\;  \{ \tilde{G}_0^{\a a},
\tilde{J}_0^{\varphi} \} = \tp_0^{\a a} .\label{commg} \ee

We wish to define an index which is nonvanishing for MSW. Consider
the following table of the R sector ground states for the center of
mass multiplet\footnote{Here we have chosen the ground state to satisfy $\tp_0^{-\pm} |0\rangle = 0$.}

\begin{table}[h]
\renewcommand{\arraystretch}{1.3}
\begin{tabular}{c|ccc|c}
state & $|0\rangle$  & $\tp_0^{+\pm} |0\rangle $& $\tp_0^{++}
\tp_0^{+-} |0\rangle $& total   \\  \hline  $(-1)^F $& 1 & - 1
$(\times 2)$ & 1 & 0 \\ \hline $ \half F^2 (-1)^F $& 0 & -1 & 2 & 1
\end{tabular}
\end{table}
From this we see that  a modified index that will not vanish due to
the trivial contribution of the center of mass multiplet can be
achieved by evaluating instead the trace of $\tilde{F}^2
(-1)^{\tilde{F}}$, as shown in the third line of the table. It is an
easy exercise to check that for states $|s \rangle $ for which all
$\tilde{G}_0^{\a b}|s \rangle \neq 0 $ the contributions from the
various members of the supermultiplet cancel, so the modified trace
defines a new index.

Interesting subtleties arise for the case in which only some of the
supercharges annihilate the state. Consider for example a state that
carries charges $q_A \in \Gamma_M$ \be |q \rangle = e^{i D^{AB} q_A
\phi_B} | 0 \rangle .\label{qsta}\ee Acting with $\tilde{G}_0^{\a
a}$ on $|q\rangle $ will produce states proportional to $\tp_0^{\a
a} | q \rangle$, as can be checked using the commutation relations
(\ref{commg}) \be (\tilde{G}_0^{\pm\pm} - p^A q_A \tp_0^{\pm\pm})
|q\rangle =0 .\ee This is simply the statement that $| q \rangle$
preserves the supersymmetries nonlinearily, as we discussed before.
As far as the index is concerned, the arithmetics is the same as for
the Ramond ground states; in particular, $\tilde{G}_0^{-a} |q
\rangle =0$ (in our random convention), and the contribution to the
modified index is again 1.

Finally, we can refine our index by introducing potentials $y^A$ for
the $h_{1,1}(M)$ conserved $U(1)$ charges. The resulting index is
called the {\it modified elliptic genus} of the MSW CFT \be
Z_{CFT}(\tau,\bar{\tau}, y^A) = Tr_R  \makebox[0.5 cm]{
$\frac{\tilde{F}^2}{2}$} \,(-1)^{\tilde{F}} q^{L_0 - \frac{c_L}{24}}
\bar{q}^{\tilde{L}_0 - \frac{c_R}{24}} e^{2 \pi i y^A q_A}
\label{ellg}\ee Here  $q = e^{2\pi i \tau}$ and the fermion number
is \be \tilde{F} = 2 J_R^3 + p^A q_A .\label{rmfn}\ee If only RM
ground states contributed to the index, then $Z_{CFT}$ would be a
holomorphic function of $\tau$, but due to the contribution of the
states of the form (\ref{qsta}), some dependence on $\bar{\tau}$ is
also introduced, since \be (\tilde{L}_0 - \frac{c_R}{24})| q \rangle
= \frac{(p^A q_A)^2}{12 D} | q \rangle .\ee Nevertheless, this
dependence is entirely due to the RM boson in the center-of-mass
multiplet, so it is under control. In fact, it can be shown
\cite{Minasian:1999qn} that the $\bar{\tau}$ dependence of the
elliptic genus is entirely of the form  \be
Z_{CFT}(\tau,\bar{\tau},y) = \sum_{\delta} Z_{\delta} (\tau)
\Theta_{\delta}(\tau,\bar{\tau},y) \label{split} \ee where
$\Theta_{\delta}$ are
 lattice theta-functions and the sum has a finite number of
 terms.

 To see how (\ref{split}) comes about, recall that the $h_{1,1}$- dimensional lattice
 lattice $\Gamma_M$ is a sublattice of the larger $6 D$-dimensional lattice $\Gamma_P$ of
 $h^{(3)}$ zero modes on the $M5$. Consider the unit
 cell $\,\mathcal{U}$ of $\Gamma_M$ in $\Gamma_P$, drawn in figure 6.
 Let $\Gamma_{M+\d}$ be a shift of the lattice $\Gamma_M$ by a lattice
 vector $\d \in \mathcal{U}$, and let $\Gamma_{\perp M + \d}$ be the lattice of vectors
 in $\Gamma_P$ normal to vectors in $\Gamma_{M+\delta}$ (this corresponds to directions
 in $\Gamma_P$ perpendicular to the plane of the figure). It is clear that one can
 rewrite the partition sum in $\Gamma_P$ as a sum over
 $\Gamma_{M+\d} \times \Gamma_{\perp M + \d}$, where $\d$ runs over the points in
 $\mathcal{U}$, a total of $det(6 D_{AB})$ of them. The sum over $\Gamma_{M+\delta}$
 gives a $\Theta$-function that encodes the $\bar{\tau}$ and $y^A$ dependence,
 while the contribution from $\Gamma_{\perp M + \d}$ is clearly holomorphic.
 We note that (\ref{split}) was used to derive an exact expression
 for the elliptic genus in some special cases \cite{Gaiotto:2006wm}.

\begin{figure}[htp]
\centering
\includegraphics[height=5 cm]{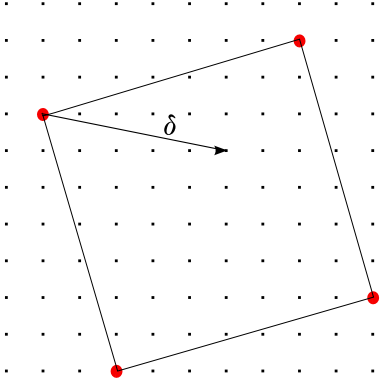}
\caption{The unit cell of the lattice $\Gamma_M$ embedded in
$\Gamma_P$. The black dots are points in $\Gamma_P$, while the red
dots are points in $\Gamma_M$. The sum over $\delta$ runs over all
points in the square, counting just one of the corners. }
\end{figure}

At last, let us mention the existence of a spectral flow
automorphism of the $\N =4$ algebra, which acts the following way on
the bosonic generators: \bea
\tilde{L}_0 &\r & \tilde{L}_0 + \eta J^3_R + \frac{c}{6} \eta^2 \non \\
J^3_R & \r & J^3_R + \frac{c}{3} \eta .\eea If the flow parameter
$\eta$ is integer the spectral flow acts on the NS and R sectors
separately, while if $\eta$ is half-integer, the NS and R sectors
get interchanged. Hence the elliptic genus (\ref{ellg}) can also be
computed as a trace over the NS sector of the appropriately shifted
variables \bea & Z_{CFT}(\tau,\bar{\tau},y^A)  = Tr_{NS} \frac{
\tilde{F}^2}{2}(-1)^{\tilde{F}} q^{L_0- \frac{c_L}{24}} \times \non
\\ &\times \; \bar{q}^{\tilde{L}_0 - \half J^3_R} e^{2\pi i y^A q_A} \eea Now
the contributions will come from the NS states that are related by
spectral flow to the Ramond ground states. These are the {\it chiral
primaries} and have $\tilde{L}_0 = \half J_R^3$. Supergravity in
$AdS_3$ (or a quotient thereof) oftentimes produces NS boundary
conditions for the fermions in the CFT living on the boundary torus.
The trace over chiral primaries is then a more natural thing to
compute from the supergravity perspective.

\bigskip
\section{The topological string }

In this lecture we briefly review what the A-model topological
string is and computes, and describe the Gromov-Witten (GW)
invariants. The many details omitted here
can be found for example
in
\cite{Pioline:2006ni,Witten:1991zz,Bershadsky:1993cx,Neitzke:2004ni,Vonk:2005yv,Marino:2004eq}.

\subsection{Twisting the string}
\medskip
The topological string is obtained by twisting an $\N=(2,2)$
superconformal $2d$ sigma-model and coupling to $2d$ gravity. We
will start with just an untwisted $(2,2)$ $\sigma$-model and build
our way up to the A-model topological string. Due to the high amount
of supersymmetry, the target space $M$ of the $\sigma$-model is
required to be K\"{a}hler. The action is then\bea
S &=& 2t \int d^2 z ( g_{i\bar{j}} \p \phi^i \bar{\p} \phi^{\bar j}  +  g_{i\bar{j}} \bar{\p} \phi^i \p \phi^{\bar j}+ i g_{i\bar{j}} \psi_+^{\bar j} \nabla_{\bar{z}} \psi_+^i \non \\
&+&  i g_{i\bar{j}} \psi_-^{\bar j} \nabla_{z} \psi_-^i  +
 R_{i\bar{j} k \bar{l}}\psi_+^i \psi_+^{\bar j} \psi_-^k
\psi_-^{\bar l} ) \non \label{ssigm}\eea where the coupling constant
$t$ sitting in front of the action can be thought of as
$\hbar^{-1}$.

 The $\N=(2,2)$ algebra has four worldsheet currents $J, G^{\pm}, T$, with spins
 $1, \frac{3}{2}, 2$ (and their antiholomorphic counterparts,  denoted by tilde).
 $T$ is the energy-momentum tensor, $J$ is the $U(1)$ $R$-symmetry current of the $\N=2$
 algebra, and $G^{\pm}$ are the conserved supercurrents for the two worldsheet supersymmetries; the $\pm$ superscript denotes their $R$-charge. From the $\N =2$ algebra, the relations that are relevant to our discussion are the R-sector anticommutators
\bea \{G^+_0, G^{+}_0 \}  & =  & \{G^{-}_0, G^{-}_0 \} = 0\\  \{G^+_0, G^-_0  \} &=& 2 (L_0 - \frac{c}{24}) \non
\eea One can combine the two possible $R$-symmetries into a vector
and an axial current $J_V = J + \tilde{J}$, $J_A = J - \tilde{J}$.
After quantising the theory, the quantum measure is invariant under
the axial $R$-symmetry only if $M$ is a Calabi-Yau manifold (of any
dimension, so far), as it must satisfy \be c_1(M) =0. \ee

Next, we would like to make our $\sigma$-model topological, which
means that the observables  in the theory should only depend on the
topological data of the target space $M$. A way to accomplish this
is by constructing a fermionic symmetry - generated by a
(scalar\footnote{The reason that we need a scalar supercharge is
that we would like our model to be defined for a worldsheet $\S$ of
arbitrary curvature and genus. If $Q$ is fermionic, then the
supersymmetry parameter $\e$ has to be a Killing spinor on $\S$,
which in general does not exist. If $Q$ is scalar though, then $\e$
is a Grassmann scalar, which exists for any worldsheet. }) BRST
charge $Q$, which is nilpotent - so that $Q^2=0$. The physical
operators in the topological theory are then defined to be $Q$
cohomology classes. Ultimately we will identify $Q$ with a
cohomology operator on $M$, such that the theory becomes
topological. We further require the energy-momentum tensor to be
$Q$-exact in order to ensure the independence of topological
correlation functions of the $2d$ worldsheet metric. All this is
accomplished by twisting according to \be T_{tws} = T - \half \p J
,\ee and taking $Q = \int G^+$. This shifts the spin of the various
operators by an amount proportional to their $R$-charge \be s_{tws}
= s - \half q .\ee In particular, now $J$ and $G^+$ have spin 1,
while $T_{tws}$ and $G^-$ have spin 2. Thus $Q = \int G^+$ can act
as our BRST charge, and $T_{tws}\sim \{ Q,G^- \}$. Next we need to
couple the $\sigma$-model to a worldsheet metric $h_{\a\b}$ and then
perform the path integral over $h_{\a\b}$ too. In order to define
string amplitudes, it is quite useful to note that the structure of
the twisted $\N=2$ algebra is isomorphic to the structure we obtain
by applying the BRST procedure to the usual bosonic string. The
correspondence between the various operators is \be (G^+, J, \,T,
G^-) \leftrightarrow (Q,J_{ghost},T, b) .\ee

Remember that if we want to compute genus $g$ amplitudes in the
bosonic string we need to integrate over the moduli space $\M_g$ of
genus $g$ Riemann surfaces, with $3g-3$ insertions of the $b$ ghost
(for $g>1$), which provide the measure. In view of the
correspondence between $G^-$ and the $b$ ghost, we define the genus
$g$ topological string amplitude as\footnote{For $g=0,1$ see
e.g.\cite{Bershadsky:1993cx,Bershadsky:1993ta,Pioline:2006ni}. Note
that the counterpart of the $\tilde{b}$ ghost could be either
$\tilde{G}^+$ or $\tilde{G}^-$, depending on the twist we perform
for the RM. Here we have made the choice corresponding to the
A-model topological string.}  \be F_g = \int_{\M_g}
\prod_{i=1}^{3g-3}\int_{\S} d^2z \,G^-_{zz} \mu_i^{ z}{}_{\bar z} \int_{\S} d^2z\,
\tilde{G}^-_{\bar{z}\bar{z}} \mu_i^{\bar{z}}{}_z \ee where $\mu_i$
are Beltrami differentials, parametrising complex structure
deformations of the moduli space of genus $g$ Riemann surfaces in
$M$. It turns out that ``ghost'' charge conservation makes almost
all $F_g$ vanish, unless we take the complex dimension of $M$ to be
three. Therefore the case in which $M$ is a  Calabi-Yau three-fold
provides the richest examples, which we will assume from now on.

The topological string ``free energy'' is defined as a perturbative
expansion in the topological string coupling constant, $g_{top}$ \be
F_{top} = \sum_{g=0}^{\infty} g_{top}^{2g-2} F_g \label{ftope}\ee
The topological string partition function is \be Z_{top} =
e^{F_{top}}\ee

Note that one could have equally considered a twist of the form $T
\r T + \half \p J$. In that case $G^-$ would become the BRST
operator. Since we can do independent twists for the LM and for the
RM, we end up with two inequivalent possible topological string
theories, depending on which supercurrent becomes the BRST charge
$$\begin{array}{ccc}
(G^+, \tilde{G}^+) &\r& \mbox{A model}  \\
(G^+,\tilde{G}^-) & \r & \mbox{B model}
\end{array}$$

\noindent As advertised, in these notes we will only concentrate on
the A-model topological string.

\subsection{The A-model topological string}

\medskip
The A-model twist is \be L_0 \r L_0 - \half J_0 , \;\;\; \tilde{L}_0
\r \tilde{L}_0 + \half \tilde{J}_0 \ee which shifts the spins as \be
S \r S - \half(J_0 + \tilde{J}_0) .\ee The local observables in the
theory are in one-to-one correspondence with the de Rham cohomology
classes on $M$. Correlation functions only receive contributions
from configurations that satisfy the classical equations of motion,
which in the A-model are  holomorphic maps from the string worldsheet to the target
space. The action can be written as a $Q$-exact term (which does
not contribute) plus
 \be \Delta S = -i t \int_{\S} \phi^*(J) =
-i t \int_{\phi(\S)} J = \,q_A t^A \label{sinst}\ee where $q_A$
are the wrapping numbers of the image of the string worldsheet
$\phi(\S)$ around the various 2-cycles dual to $\omega_A$ and
 $t^A$ are the
complexified K\"{a}hler moduli of the Calabi-Yau . We also set $t =
(\a')^{-1}$. To compute the topological string partition function,
we need to integrate over the moduli space $\M_{g,{\bf q}}$ of maps
from a genus $g$ Riemann surface to the Calabi-Yau $M$ - where the
image belongs to the homology class ${\bf q}$ of $M$, weigh it by
the exponential of minus the euclidean action (\ref{sinst}), and
then sum over genera. From (\ref{sinst}) it is clear that the
A-model topological string only depends on half the information of
the Calabi-Yau - that is - only on the K\"{a}hler structure.

The {\it Gromov-Witten} invariants $d_{g,\, q_A}$  are
defined as the expansion coefficients in \be F_{GW} (t^A) = \sum_{g=0}^{\infty}
g_{top}^{2g-2} \sum_{q_A} d_{g, \,q_A} e^{- q_A t^A} \ee which
encode the contributions which are nonperturbative in $\a'$.
$d_{g,\, q_A}$ is roughly the euler character of $\M_{g, \, q_A}$.

There are also perturbative contributions to $F_{top}$, that are not
encoded in $F_{GW}$. These contributions can be computed in just
low-energy field theory, and are only present at genus 0 and 1 \be
F_{pert} (t^A) = -i \frac{(2\pi)^3}{g_{top}^2} D_{ABC} t^A t^B t^C -
\frac{i\pi}{12} c_{2A}t^A \label{fprt}. \ee

\bigskip
$\bullet${\bf What the topological string computes}

\medskip
Recall from our discussion in section (2.7) that higher curvature
corrections to the $\N=2$ supergravity action in $4d$ are encoded in
a holomorphic function of the vector moduli and the Weyl multiplet
\bea S_{corr}& \sim & \int d^4 x d^4 \theta \F_g(X^{\L}) W^{2g} +
c.c.
\\  \label{actcomp} & \sim & \int d^4 x [ F_g (t) R_-^2 T_-^{2g-2} +
F_{g}(\bar{t}) R_+^2 T_+^{2g-2}] \non \eea where $T_{\pm}$ and
$R_{\pm}$ are the self-dual and anti-self-dual parts of the
graviphoton and Weyl tensor respectively. In usual string theory,
such a contribution to the effective action can be computed by
evaluating the correlation function of two graviton and $2g-2$
graviphoton vertex operators on a genus $g$ string worldsheet. This
computation has been performed in \cite{Antoniadis:1993ze} and it
was found that the amplitude for this scattering process is
precisely given by the genus $g$ topological string amplitude $F_g$.
\begin{figure}[htp]
\centering
\includegraphics[height= 4 cm]{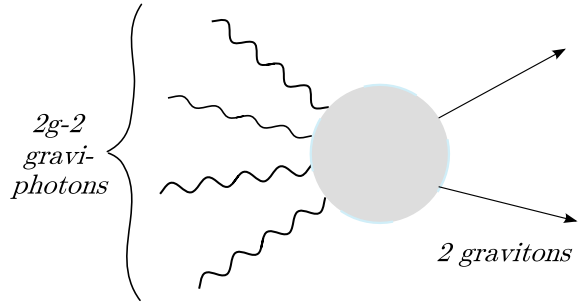}
\caption{Scattering of two gravitons and of $2g-2$ graviphotons
gives $F_g$.}\label{ads2}
\end{figure}

\section{The OSV conjecture}
\medskip
In lecture 2 we discussed black hole attractors in the context of
type IIA string theory compactified on a Calabi-Yau $M$. We found
that the entropy of these black holes only depended on the
K\"{a}hler moduli of the compactification, which in turn are fixed
to their attractor values at the horizon of the black hole. Also,
remember that hypermultiplet scalars - which in type IIA correspond
to the complex structure deformations of $M$ - completely dropped
out of the story. Interestingly, in the A-model topological string
half of the information on the Calabi-Yau data also drops out due to
the twist.
 This is a first clue towards a connection between
attractor black holes and topological strings: they are both
functions on half of the Calabi-Yau moduli space.  It would be
strange if in string theory we had two natural functions of the same
variables that were not related in some simple way. Indeed we shall
see that, when the comparison between the two quantities is properly
formulated, the relation between the indexed black hole entropy and
topological string partition functions appears to take the simplest
imaginable form.

\subsection{Mixed partition functions}
\medskip
If we want to make an exact or all orders comparison between the
partition functions of two systems, we must first be precise about
exactly which ensemble we are using. A statistical system or
ensemble can be characterized either by conserved quantities such as
the energy $E$ or by conjugate potentials such as the inverse
temperature $\beta$. There are associated microcanonical and
canonical partition functions $\Omega$ and $Z$, , which for the case
of energy and temperature are related by \be Z(\beta)=\sum_E
e^{-\beta E}\Omega(E).\ee More generally, if there are many
conserved charges/potentials, partition functions can be defined in
which some variables are treated canonically and others
microcanonically. We refer to these as {\it mixed} partition
functions associated to {\it mixed} ensembles.

For every type of (mixed) ensemble there is an associated definition
of entropy. For example, in the microcanonical ensemble the entropy
is defined as \be\label{me} S_E(E)=\ln \Omega (E),\ee  while in the
canonical ensemble it is defined by \be\label{ce}S_\beta(\beta)= \ln
Z-\beta
\partial_\beta\ln Z.\ee   To leading order in the saddle point
approximation one has \be S_\beta(\beta(E))\sim S_E(E)\label{fdv}\ee
with $ \beta(E)\equiv  \partial_ES_E$. However in general, for a
finite system and beyond leading order, $S_\beta$ and $S_E$ are not
related as simply as in (\ref{fdv}) - by just a change of variables.
Rather, an integral transform is needed. Hence if we want to discuss
subleading corrections to the entropy we must specify exactly which
ensemble we are using to define the entropy. It is meaningless to
give an all-orders formula for the entropy without this
specification.

So the question arises: exactly which entropy does Wald's formula,
discussed in section 2.7.2,  compute? Wald's derivation is
ultimately based on a quantum field theory analysis. The answer then
follows from the fact that in quantum field theory the boundary
conditions amount to an implicit choice of ensemble. For example, if
we sum over geometries with asymptotic periodicity $\beta$, we are
working in the canonical fixed-temperature ensemble. There is no
field theory path integral formulation for the microcanonical
ensemble. The best we can do is a Laplace transform \be \Omega(E)=
\int d\b \,e^{\b E} Z(\b) ,\ee which requires knowledge of $Z(\b)$
for all $\b$. So for a finite-temperature black hole  Wald's formula
computes $S_\b$ rather than $S_E$.

What about charges? For electromagnetic charges, instead of fixing
the radius of the circle at infinity, in the field theory path
integral one fixes the boundary value of the field $A_0=\phi$ (the
electric Wilson line) at infinity, as well as the topological class
of the gauge field. This corresponds to fixing the magnetic charge
$q_m$ while summing over all electric charges with weights $e^{-\phi
q_e}$. Hence field theory gives a mixed ensemble which treats
electric charges canonically and magnetic charges microcanonically.
This then is the ensemble in which the entropy computed by Wald's
formula is defined\footnote{See also \cite{LopesCardoso:2006bg} for
a discussion of thermoodynamic ensembles.}. We will see this is
crucial for comparison of Wald's formula for the entropy to the
topological string partition function.

We note that an independent but similar issue is whether or not the
the ensemble is weighted with minus signs for fermions. This is
again related to path integral boundary conditions and is discussed
below.

\subsection{Brute force derivation of OSV}
\medskip

Now we would like to get to work and compute the corrections to the
entropy, using Wald's formula. The first question is, which terms
from the corrected $\N=2$ action do actually contribute. The working
assumption has been \cite{LopesCardoso:1998wt}\footnote{Reference
\cite{LopesCardoso:1998wt} does not address all the issued involved
and this point remains in need of further clarification. Some
relevant, but still incomplete, observations are made section 4.3.1
below.} that if we compute the indexed entropy - defined in the
usual way as a Legendre transform of the supersymmetry-protected
indexed partition function (more precisely, the modified elliptic
genus (\ref{ellg})) that there is a (perturbative)
nonrenormalization theorem implying that it receives contributions
only from supersymmetry-protected terms in the effective action.
These are the terms which involve integrals over only a chiral half
of superspace and so cannot involve hypermultiplets. Hence we expect
that only terms in the action that are built exclusively out of
vector multiplets can correct the entropy. These are given of course
by (\ref{actcomp}), and their contribution to the entropy has been
quoted in section 2.7 \be S_{BH} = \frac{\pi i}{2} (q_{\L} \bar{C}
\bar{X}^{\L} - p^{\L}\bar{C} \bar{F}_{\L}) - \frac{\pi}{2} Im (C^2 W
\p_W F). \label{bhet}\ee Naively, one might evaluate the $X^\L,F_\L$
at their attractor values (\ref{genatt}) and (\ref{watt}) and view
this as an expression for the entropy as a function of the magnetric
and electic charges $p^\L$ and $q_\L$. However, according to the
discussion of the previous section, this is not correct:
(\ref{bhet}) comes from a mixed ensemble and should be viewed as a
function of the magnetic charges $p^\L=Re CX^\L$ and the electric
potentials, which shall be identified shortly as  $\phi^\L= \pi Im
CX^\L$.

 The above expression for the entropy
 can be rewritten in a nicer form  by using the homogeneity property of the prepotential\footnote{$2F = X^{\L} F_{\L} + W \p_{W} F$.} and the attractor equations
\be S_{BH}(p,\phi) = - \pi Im F(CX^{\L},256) + \pi Im (CX^{\L})\,
q_{\L} ,\ee which let us eliminate $F$ and
$F_\L$ in favor of $p^\L$ and $\phi^\L$. In terms of $\phi^{\L}$ and
the imaginary part of $F$ this is \be S_{BH} = \F(p^{\L},\phi^{\L})
+ \phi^{\L} \, q_{\L} ,\ee where \be \F(p,\phi) = - \pi Im F(p^{\L} +
\frac{i}{\pi} \phi^{\L}, 256) .\ee The second half of the attractor
equations then reads \be q_{\L} = -\frac{\p}{\p \phi^{\L}}\F(p,\phi)
\ee which implies that $S_{BH}$ is obtained from $\F(p,\phi)$ in
exactly the same way that the entropy is obtained from the logarithm of
the mixed partition function. Namely, $S_{BH}$ can be thought of as the Legendre transform with
respect to the canonical variables only \be S_{BH}(q,p) = \F(\phi,p)
- \phi^{\L} \frac{\p}{\p \phi^{\L}} \F(\phi,p) \ee provided that the
$\phi^{\L}$ are indeed identified as the conjugate potentials to the
electric charges $q_{\L}$.

 In conclusion, $\F(p,\phi)$ is the logarithm of the partition function computed in a mixed
ensemble, in which one fixes the magnetic charges and electric
potentials at infinity. We will loosely refer to $\F$ as the free
energy though strictly speaking it differs by a factor of $-\beta$
from the usual definition.

One relatively illuminating way to write the mixed free energy is by
picking a gauge in which $C=2Q$, where $Q$ is the graviphoton
charge. Then, using the homogeneity properties of $F$, we have \be
\F = -4 \pi Q^2 Im \left[\sum_g F_g \left( \frac{p^{\L}+ i
\phi^{\L}/\pi}{2 Q}\right) \left(\frac{8}{Q} \right)^{2g} \right]
\ee which shows very clearly the perturbative nature of $Z_{BH}$ as
an expansion around large graviphoton charge. Now, the topological
string partition function is also defined by a perturbative
expansion in $g_{top}$ (\ref{ftope}) and, as mentioned in section
2.7, it is proportional to the supergravity prepotential $F(CX^{\L},
256)$. By comparing the first two terms (\ref{fprt}) in the
expansions of the two prepotentials, one can fix all normalizations
and finds \be F(CX^{\L}) = - \frac{2i}{\pi} F_{top} (t^A,g_{top})
\label{ftopnm} \ee with the following correspondence between the
arguments on the two sides of the equation \footnote{For a D4-D2-D0
black hole, in the large-charge, large Calabi-Yau volume
approximation, $g_{top} = \frac{4 \pi^2}{\phi^0} \sim
\sqrt{\frac{q_0}{D}} << 1$ if we uniformly scale up the charges of
the black hole. This is consistent with the fact that we are
performing a perturbative expansion in the topological string
coupling constant. } \be CX^\L=p^\L + i \phi^\L/\pi,\;\;t^A =
\frac{CX^A}{CX^0}, \;\; g_{top} = \pm \frac{4 \pi i}{CX^0}
\label{reltps} \ee From (\ref{ftopnm}) one finds \be \ln
Z_{BH} = - \pi Im F = 2 Re F_{top} \ee which can be
written in the more expressive form \be Z_{BH}(\phi^{\L},p^{\L}) =
|Z_{top}(t^A,g_{top})|^2 \label{osvc}\ee with the variables
identified as in (\ref{reltps}). As there are several assumptions
that went in to the derivation of (\ref{osvc}) it is known as the
OSV conjecture. It is important to note that the conjecture is a
statement about the equality of two perturbation expansions. Indeed
at the nonperturbative level it is not clear how either side of
(\ref{osvc}) is even defined.

Note that the factor of `2' in the exponent of (\ref{osvc}) was
obtained by the brute force method -- just carefully keeping track
of all normalizations. So far there is no hint as to why it should
not equal 1 or 17 or any other real number. Such a simple relation
demands a simple physical explanation, which we will provide in the
next section.

\subsection{Why $Z_{BH}=|Z_{top}|^2$}

\medskip
In this section we will give a heuristic explanation of why the two
complex conjugate factors $Z_{top}$ and $\bar Z_{top}$ appear in the
black hole partition function. The goal here is not to give a
perturbative proof of the OSV conjecture -- many points would have
to be filled in and clarified -- but rather to provide a compelling
physical picture\footnote{We follow here \cite{Beasley:2006us},
which in turn built on earlier discussions in
\cite{deBoer:2006vg,Kraus:2006nb,Gaiotto:2006ns,Denef:2007vg}.}.

The basic idea is simply to evaluate the black hole partition
function in a string perturbation expansion around a euclidean
saddle point, which is a euclideanized attractor geometry (and then
compare it to the perturbation expansion for the topological
string). It is then argued that at $g$ loops, string perturbation
theory is saturated by genus $g$ worldsheet (anti)instantons  which
wrap (anti)holomorphic cycles in the Calabi-Yau and localize to the
(south) north pole of the horizon $S^2$. This is possible due to the
peculiarities of the $AdS_2\times S^2$ supersymmetries, which are
broken in exactly the same way for an instanton at the north pole
and an anti-instatnon at the south pole. The two factors in $Z_{BH}$
then come form the instanton sum at the north pole and and the
anti-instanton sum at the south pole.

\subsubsection{$M$-theory lift}
\medskip

Our starting point will be the IIA string partition function
-denoted $Z_{IIA}$ - on the euclidean attractor geometry. As will be
seen explicitly below, the sum over worldsheets gives a finite
non-zero answer. Furthermore, there must be  many supersymmetric
cancellations, as in the limit in which the  $AdS_2\times S^2$
radius goes to infinity the term which scales as volume is the
cosmological constant and must vanish.

So the question arises - what is $Z_{IIA}$ computing? Since the
computation involves a choice of attractor geometry, it should be
related to the associated black hole. Since it is supersymmetry
protected, it should be some supersymmetry-protected invariant
associated with the black hole. We know of only one such object--the
modified elliptic genus. So the obvious guess is \be
Z_{IIA}=Z_{CFT}+{\cal O}(e^{-\frac {1}{g}}).\label{iiabh}\ee Note
that since the left hand side is defined only in string perturbation
theory, this is at best a perturbative relation (at large charges).
If one further assumes $Z_{BH}=Z_{CFT}+{\cal O}(e^{-\frac {1}{g}})$
one has $Z_{IIA}=Z_{BH}$.

There are a number of ways one might go about demonstrating
(\ref{iiabh}). The most straightforward approach would be to refine
and adapt the methods of \cite{Banados:1993qp,Callan:1994py}. The
subtleties would be to carefully understand the euclideanization,
especially of the RR fields which become complex, and the fermion
boundary conditions. Instead of this more direct method,
\cite{Beasley:2006us} adopted a shortcut involving a lift to
M-theory. This had the disadvantage of using a nonperturbative
relation to demonstrate a perturbative one, but on the other hand
the construction, which we now review, is illuminating in its own
right.

 To keep the equations uncluttered, we just consider the
 euclideanised near-horizon geometry of a D4-D0 black hole.
 The metric on the $AdS_2 \times S^2$ part is
\be ds_4^2 = 4 l^2 \left( \frac{dr^2 + r^2 d\th^2}{(1-r^2)^2} +
\frac{1}{4} d \Omega_2^2 \right) ,\ee the radii of the $AdS_2$ and
$S^2$ are equal to \be l=  \frac{g_s \phi^0}{\pi} \sqrt{\a'} = g_s
\sqrt{ \frac{D \a'}{q_0}}  + \mbox{corrections} \label{adsrad} \ee
There is also a graviphoton field \be A^{(1)} = - \frac{2 i
\phi^0}{\pi} \frac{d \th}{1 - r^2} .\ee Note that this potential is
purely imaginary - as expected when we continue electric fields to
euclidean space. Noting that the graviphoton is just the connection
on the M-theory circle (parametrized here by $x^{11}$), we simply
end up with the M-theory metric \bea &ds_{11}^2 = g_s^2 \a' \left(
dx^{11} - \frac{2i \phi^0}{\pi} \frac{d\th
}{1-r^2} \right)^2 + 4 l^2\frac{dr^2 + r^2 d\th^2}{(1-r^2)^2} \non \\
&+ l^2 d\Omega_2^2 + ds_{CY}^2 .\eea We immediately recognise an
$AdS_3$ factor in the first line of the above equation. Note that
some of the metric components are imaginary, so this is a {\it
complexified} $AdS_3$ quotient. The interpretation of this
complexified geometry  can be gleaned from looking at the torus that
is located at the boundary $r=1$. Keeping the leading terms in the
metric as we take $\e = 1-r \r 0$, we find the conformal metric on
the boundary \be ds^2_{bnd} = d\th^2 + \frac{2\pi i}{\phi^0}\, d\th
\, d x^{11} + \Q(\e) .\label{bndm}\ee This can be put in the
standard form \be ds_{\tau}^2 = |d\th + \tau dx^{11}|^2 \ee if we
take the modulus of the torus parametrised by $(\theta, x^{11})$ to
satisfy \be |\tau|^2 = (Re\tau)^2 + (Im\tau)^2 = 0 \; \mbox{and} \;
Re\tau = i \frac{\pi}{\phi^0} .\label{reim} \ee Now, according to
$AdS_3/CFT_2$, the partition function on this euclidean M-geometry
is a partition function of the MSW CFT on the boundary torus
(\ref{bndm}) \bea Z_M(\tau) &\sim & Tr_{MSW}  e^{2\pi i Re \tau (L_0
- \tilde{L}_0)} e^{- 2\pi Im \tau (L_0 + \tilde{L}_0)} \non
\\ &\sim & Tr_{MSW} e^{-\frac{4\pi^2}{\phi^0} L_0} .\label{wht} \eea The
main point we wish to stress here is that $\tau$ appears as a
weighting factor for the left movers only. This result might have
been anticipated from the fact that the lorentzian D4-D0 black hole
is dual to a state in the MSW CFT in which the left movers are
thermally excited but the right movers are in their ground state.

Consideration of the fermion boundary conditions
\cite{Beasley:2006us} indicates that, after a spectral flow one
obtains (\ref{wht}) with the trace in the R sector and an insertion
of $(-)^F$. An ${F}^2$ insertion is needed when the degrees of
freedom corresponding to the center-of-mass multiplet (ignored so
far) are included: without the insertion this multiplet gives a zero
prefactor. In the bulk $AdS$ description these appear as boundary
``singleton'' modes. Exactly how this all works out has not been
carefully analyzed and the $F^2$ insertion is accordingly suppressed
in the following.

Of course, the index is defined for fixed electric potentials on the
boundaries, so we need to add D2-brane charges, and then sum over
them. Upon reinstating the D2 potentials $\phi^A$, the partition function
we are computing becomes  \be
 Z_{IIA} (\phi^0,\phi^A, p^A) \sim Tr (-1)^F e^{-\frac{4\pi^2}{\phi^0}L_0 - q_A \frac{\phi^A}{\phi^0} } \label{trnsz}
,\ee where here and elsewhere ${\cal O}(e^{-1/g})$ corrections are
implicit.

In order to make the connection with the topological string, it is
useful to rewrite (\ref{trnsz}) in terms of topological string
variables. Using  (\ref{reltps}) for the D4-D2-D0 black hole, we can
rewrite (\ref{trnsz}) as \bea
Z_{IIA} &=& Tr (-1)^F e^{-g_{top}L_0 - q_A (t^A + i \pi p^A)} \non \\
&=& Tr (-1)^{\tilde{F}} e^{-g_{top} L_0 - q_A t^A} \eea where
$\tilde{F}$ is just the modified RM fermion number (\ref{rmfn})
needed for modular invariance and $q_A t^A$ is the correct string
instanton action.

In conclusion, string theory on the euclidean attractor geometry is
expected to give a perturbative expansion of the supersymmetric
partition function of the associated black hole.

\subsubsection{Computing $Z_{IIA}$ }
\medskip

Now we must evaluate the perturbative string loop expansion of the
type IIA partition function on the attractor geometry. This
computation could be set up in the NS-R, Green-Schwarz or hybrid
formalism, each of which has its own complementary set of advantages
and disadvantages. In the NS-R formalism the connection to the
topological string is clearest, but RR fluxes are hard to deal with.
In the Green-Schwarz formalism the action in RR backgrounds is
known, but the reduction to the topological string is only partially
worked out \cite{Beasley:2005iu}. The hybrid formalism, as it more
or less treats the $4d$ spacetime part in the Green-Schwarz language
and the Calabi-Yau part in the NS-R language is perhaps ultimately
the most suitable, but unfortunately it is less developed at
present.

In \cite{Beasley:2006us} we have used the Green-Schwarz formalism.
The worldsheet action effectively splits into an internal $CY_3$
term and an external $AdS_2\times S^2$ piece, and the computation
factorizes. The details of exactly how the internal piece reproduces
the topological string have not all been completed, but the presumed
equivalence of the Green-Schwarz and NS-R strings imply they must
work out, as shall be assumed herein. Although of interest in their
own right these details are essentially the same in $CY_3\times
\mathbb{R}^4$ and in an attractor geometry and are not the focus of
our current investigation. Here we focus on the $AdS_2\times S^2$
factor where interesting new features arise.

\begin{figure}[h]
\centering
\epsffile[2 22 200 265]{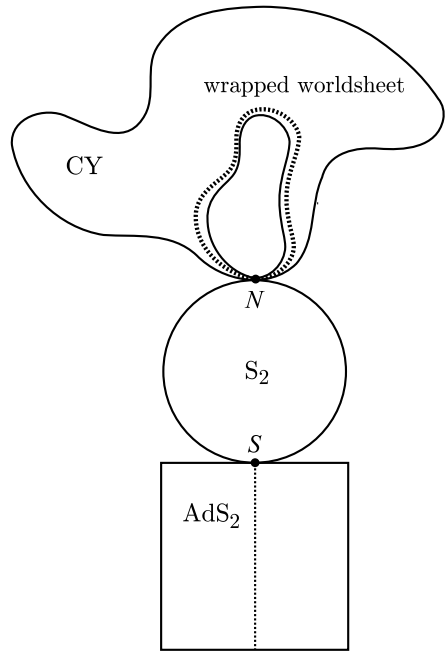}
\caption{Illustration of a contributing worldsheet instanton, which
wraps a curve in the Calabi-Yau, sits at the north pole of the $S^2$
and the center of $AdS_2$. Anti-instantons wrap the same curve but
with the opposite orientation, sit at the south pole and the center
of $AdS_2$.}\label{instanton}
\end{figure}

Let us now summarize the key steps in the somewhat technical
computation \cite{Beasley:2006us}. The classically supersymmetric
genus $g$ worldsheets $\S_g$ wrap holomorphic cycles (instantons) or
antiholomorphic cycles (anti-instantons) in $M$ and sit at any point
in $AdS_2\times S^2$. However, as will be important momentarily,
which supersymmetry is preserved depends on the point chosen in
$AdS_2\times S^2$. The worldsheet sum is organized as an expansion
about these configurations. The internal part (presumably) gives a
factor of $F_g(CX^\L)$, as we will try to sketch below. Afterwards,
we are just left with the ordinary integral of a (-1)-brane
zero-mode action in $AdS_2\times S^2$.

 The ``fields''
consist of  the four spacetime coordinates $X^\mu$, their four
chiral goldstino superpartners $\theta_{1,2}^{\a}$, $4g$ fermionic
fields $\rho^\a_{i}$ coming from the $4g$ zero modes of the
canonically conjugate momenta to the $\th_{1,2}^{\a}$ on $\S_g$, and
possibly some antichiral fields $\chi^{m\dot{\a}}$, $\chi^{{\bar m}
\dot{\a}}$. The latter correspond to zero modes of the fermionic
superpartners of the normal fluctuations of the worldsheet $\S_g$
inside the Calabi-Yau, which occur whenever $\S_g$ is not an
isolated curve in $M$.

 There is no action for
$X^\mu$ or $\theta_{1,2}$. The action for $\rho^\a$ is simply
\cite{Beasley:2005iu} \be S_{int} = \int W_{\a\b} \rho^{\a} \wedge
\rho^{\beta} ,\ee $W_{\a\b}$ is the anti-self-dual part of the
graviphoton field strength, with attractor value $g_{top}$. Thus the
$\rho$ contribution to the path integral is \be
 \int d^{4g} \rho\, e^{- \int  W \rho^2} =
 \int d^{4g}\rho (W \rho^2)^{2g} = g_{top}^{2 g} .\ee

The fermionic zero-modes $\chi^{m,{\bar m}}$ and their bosonic
superpartners are described by supersymmetric quantum mechanics on
the moduli space $\M_{g,q_A}$ of holomorphic deformations of the curve inside
$M$. The supersymmetric index we are computing is
independent of the details of this quantum mechanics and is known to
equal the euler character\footnote{A typical four-supercharge action
contains a coupling $ \int R_{m\bar{n} p\bar{q}}\, \chi^{m\dot{\a}}
\chi^{\bar{n} \dot{\b}} \chi^p{}_{\dot{\a}} \chi^{\bar
q}{}_{\dot{\b}} $, and one can bring down powers of this interaction
to make the path integral over the $\chi^m$ zero modes not vanish.
One can easily check that this procedure yields the euler character
of moduli space. } of  $\M_{g,q_A}$, which is roughly the
Gromov-Witten invariant $d_{g,q_A}$. Combining it with the factor
that comes from the instanton action, $e^{-q_A t^A}$, we finally
obtain $F_{g,\,q_A}(t^A)$.

It remains to consider the integral over the position zero-mode
$X^\mu$ and the goldstinos $\theta_{1,2}$. The former is
proportional to the volume of $AdS_2\times S^2$ while the latter
vanishes because there is no action. Hence \be F_4 = \int d^4 x\,
d^2 \th_1 d^2 \th_2 e^{-0}= \infty \times 0 = ?! \ee This integral
can be defined by use of localization. One adds an exact term
$\Delta S = \delta K$ to the action with an arbitrary real
coefficient $t$ \be F_4 = \int d^4 x\, d^2 \th_1 d^2 \th_2
e^{-t\Delta S}.\ee
 $\delta$ here is a nilpotent combination of the kappa-symmetries
and supersymmetries and $K$ is a judiciously chosen operator.
Exactness and nilpotency imply that the integral is independent of
$t$.  It is most easily evaluated at $t \to \infty$. The
contributions will then be localized to $\delta$-invariant
configurations, which are instantons and anti-instantons sitting at
the center of $AdS_2$ and the north pole and respectively south pole
of the sphere. The answer we get is some nonzero constant $C$, whose
value is most easily determined by comparison to supergravity.

Now let us put everything together. Concentrating for now just on
the contribution of the instanton at the north pole, we have the
following
\begin{description}
\item - a genus-independent constant factor C coming from the integral over the bosonic and goldstino zero-modes
\item - a factor of $F_g(t^A)$ coming from the internal worldsheet partition function.
\item - a factor of $g_{top}^{2g}$ from the zero-mode integrals for $\rho^\a$.
\end{description}
Summing these over genera we obtain \be \sum_{g} C F_g \,
g_{top}^{2g} = g_{top}^2 C \,F_{top} .\ee The anti-instanton sitting
at the south pole of the sphere similarly gives a factor of
$g_{top}^2 \bar{C} \bar{F}_{top}$. Exponentiating the sum of the two
contributions we almost have the OSV relation, except we still need
to find the value of $C$. This can be fixed by comparing any term in
the expansion of $Z_{BH}$ with the corresponding term in the
topological string expansion, as was done in section 3.6, and is $C
= g_{top}^{-2}$. In conclusion this indicates that  in perturbation
theory

\begin{center}
\boxed{ Z_{BH} = Z_{IIA}= |Z_{top}|^2 }
\end{center}
which is the relation we set out to explain.

\section{$\bf Z_{BH},~~Z_{CFT},~~Z_{IIA},~~Z_{top}$ and all that}

In these lectures several closely related partition functions, all
denoted with the letter $"{\bf Z}"$, have appeared. We will close by
summarizing what we mean by these  objects, how they are computed
(or not) and how they are related.

\subsection{Definitions}
\medskip

\noindent$\bullet \bf Z_{BH}$  This partition function is deduced
from a perturbative spacetime analysis of the thermodynamic
properties of black holes, which at leading order are governed by
the famous Bekenstein-Hawking area-entropy law. Subleading
corrections are computed using Wald's formula. We are not using this
symbol herein to denote an object defined nonperturbatively by
counting BPS states.

  In our case, we are interested in extremal black holes, and the
  corrections to the area law are computed in a power series in the
  inverse graviphoton charge $Q$, which governs the size of the
  black hole. Moreover we are interested in a supersymmetric partition function
  with (among other things) a $(-)^F$ insertion for fermions.
  In principle the effect of this on the spacetime analysis and
  Wald's formula should be derivable from first principles, but in
  practice no one has done so. The arguably well-motivated
  assumption has been that the insertions are accounted for by including only
  $F$ term corrections to Wald's formula.

  Any definition of $Z_{BH}$ along these lines will necessarily entail some
  dependence on the moduli of the Calabi-Yau compactification in which the
  black hole sits. This dependence has a beautiful and intricate mathematical
  structure following from the ``split attractor'' phenomena, the unraveling of which has been the subject of recent
  progress \cite{Denef:2007vg}.

  This definition of $Z_{BH}$ is essentially perturbative in
  $\frac{1}{Q}$, and
  does not have any obvious non-perturbative completion.

\medskip

\noindent$\bullet \bf Z_{CFT}$  This is defined as the modified
elliptic genus, with appropriate potentials, of the MSW CFT. In
principle this has a chance of being defined nonperturbatively.
However there are a number of complicating issues which have not
been understood (see e.g. \cite{Minasian:1999qn}). An important one is that
one must regulate IR divergences associated with noncompact Coulomb
branches - where the M5-branes can separate. Related issues are the
holomorphic anomaly, background dependence and singularities in the
moduli space of divisors. This is the only potentially
nonperturbatively defined object on our list.

\medskip

\noindent$\bullet \bf Z_{IIA}$ This is the IIA partition function
computed in the string genus expansion around a euclidean attractor
geometry. This expansion is formally equivalent to an expansion in
the RR graviphoton field, so the problems of dealing with background
RR fields are not so severe here. There are the usual issues
associated with euclidean quantum gravity and complex saddle points
that must be dealt with.

\medskip

\noindent$\bullet \bf Z_{top}$ This is the topological string
partition function, and is defined in the worldsheet genus
expansion. The expansion coefficients have a precise mathematical
definition in terms of the Gromov-Witten invariants associated to
maps of a genus $g$ Riemann surface in to a Calabi-Yau space. So
$Z_{top}$, unlike $Z_{IIA}$ and $Z_{BH}$, is fully defined in
perturbation theory. No nonperturbative definition of the full
function $Z_{top}$ is known.

\medskip

\noindent$\bullet \bf \Omega(p,q; \tau_{\infty})$ Another
interesting object, not directly discussed in these lectures but
ultimately very relevant is $\Omega (p,q; \tau_{\infty})$. This is
the (weighted) number of BPS states with charges $(p,q)$ which
depends (due to jumping phenomena) on the asymptotic value
$\tau_{\infty}$ of the Calabi-Yau moduli, and should be
nonperturbatively well defined. A recent and mathematically precise
definiton of $\Omega$, a derivation of some of its properties and a
corresponding precise version of the strong $OSV$ conjecture can be
found in \cite{Denef:2007vg}.  In our list of objects above, only
$Z_{CFT}$ is non-perturbatively defined and so potentially related
to $\Omega$.

\subsection{Relations}
\medskip

At a very formal level, $Z_{BH}=Z_{IIA}=Z_{CFT}=|Z_{top}|^2$, in the
appropriate perturbation expansions and with the appropriate
identifications of parameters. Beyond perturbation theory, these
relations are not even formally true. In practice, none of these
equalities have been proven or even stated herein with total
precision. Let us now comment on the various equalities.
\medskip

\noindent$\bullet \bf Z_{BH}=Z_{CFT}$

This is a deep statement about the holographic nature of string
theory, and was the early form of $AdS/CFT$ duality \cite{Strominger:1996sh}. At
leading order it has been tested in many examples. Some tests of
subleading terms are discussed in \cite{LopesCardoso:1999cv,Maldacena:1997de}. Beyond
subleading order a better understanding of the definition of the
left hand side is needed.

\medskip

\noindent$\bullet \bf Z_{BH}=|Z_{top}|^2$

This is the weak form of the OSV conjecture. It equates terms in two
differently defined perturbation expansions
\cite{Gaiotto:2006wm,Gaiotto:2006ns,Aganagic:2004js,Aganagic:2006je,Dabholkar:2004yr,Dabholkar:2005dt,Verlinde:2004ck,Shih:2005he,Jafferis:2005jd}.
One of the main issues here is understanding how the background
dependence works out. In essence however it is a statement about
string perturbation theory and does not have the dynamical depth of
the preceding equality. It could in principle be precisely stated
and proved without any understanding of non-perturbative string
theory.

\medskip

\noindent$\bullet \bf Z_{BH}=Z_{IIA}$

This asserts that the black hole partition function can be
perturbatively repackaged as a euclidean calculation on the
attractor geometry. It is a stringy version of the black hole
methods of semiclassical euclidean quantum gravity, adapted to deal
with an index. Arguments for this equality were given in
\cite{Beasley:2006us}. Assuming that this and the preceding equality
is valid for the same definition of $Z_{BH}$ leads to
$Z_{IIA}=|Z_{top}|^2$, along with a physical explanation of the two
complex conjugate factors on the RHS of the OSV relation.

\medskip

\noindent$\bullet \bf Z_{CFT}=|Z_{top}|^2$

This is the strong form of the OSV conjecture, and comes from
combining holographic duality with the perturbative observations of
\cite{Ooguri:2004zv}. Since the right hand side is defined only
perturbatively, it is at most a perturbative statement. Still, it is
extremely interesting in part because both sides are potentially
rigorously defined mathematically. From the mathematical point of
view, it is a totally unexpected relationship between moduli spaces
of divisors and maps of curves into a Calabi-Yau space.
   In this paper we have argued from the stringy perspective that a
   relation of this form is to be expected. It would be a daunting
   task to turn these arguments in to a rigorous mathematical
   derivation. While these arguments are informative, in the end
   this relation should be regarded as a conjecture to be tested by
   direct computation.
\bigskip

In conclusion, the conjectured OSV relationship between the
all-orders expression for the entropy of large BPS black holes and
the all-orders expression for the topological string partition
function potentially allows for precision tests of non-perturbative
string theory. We hope to have provided the reader with a flavor of
this exciting subject that melts together black holes, attractors,
and topological strings, ans at the same time raises interesting new
challenges and puzzles.

\bigskip $\bullet$ {\bf Acknowledgements} \medskip

We are grateful to F. Denef, D. Gaiotto, G. Moore and Xi Yin for
very helpful conversations. Special thanks to C. Garc\'{i}a for help
with the figures. Finally we wish to thank Laurent Baulieu, Pierre
Vanhove, Paul Windey, Mike Douglas, Jan de Boer and Eliezer
Rabinovici for a really stimulating and fun school. This work has
been partially supported by DOE grant DE-FG02-91ER40654.

\bibliographystyle{utcaps}
\bibliography{cargese}

\end{document}